\documentclass[]{aa} 
\usepackage{amsmath}
\usepackage{graphicx}
\usepackage{graphics}
\usepackage{subfigure}
\usepackage{txfonts}
\usepackage{natbib}
\bibpunct{(}{)}{;}{a}{}{,}

\def\teff{\ifmmode T_{\rm eff} \else $T_{\mathrm{eff}}$\fi}

\def\ltsima{$\buildrel<\over\sim$}
\def\lsim{\lower.5ex\hbox{\ltsima}}

\newcommand{\hii}{H~{\sc ii}}
\newcommand{\ha}{\ifmmode {\rm H}\alpha \else H$\alpha$\fi}
\newcommand{\hb}{\ifmmode {\rm H}\beta \else H$\beta$\fi}
\newcommand{\lya}{\ifmmode {\rm Ly}\alpha \else Ly$\alpha$\fi}

\newcommand{\ebv}{\ifmmode E_{\rm B-V} \else $E_{\rm B-V}$\fi}
\newcommand{\av}{\ifmmode A_{\rm V} \else $A_{\rm V}$\fi}
\def\micron{$\mu$m}

\def\ergscm{erg s$^{-1}$ cm$^{-2}$}
\def\msun{\ifmmode M_{\odot} \else M$_{\odot}$\fi}
\def\msunyr{\ifmmode M_{\odot} {\rm yr}^{-1} \else M$_{\odot}$ yr$^{-1}$\fi}
\def\zsun{\ifmmode Z_{\odot} \else Z$_{\odot}$\fi}
\def\lsun{\ifmmode L_{\odot} \else L$_{\odot}$\fi}

\def\mup{\ifmmode M_{\rm up} \else M$_{\rm up}$\fi}
\def\mlow{\ifmmode M_{\rm low} \else M$_{\rm low}$\fi}

\def\bacs{B$_{\rm 435}$}
\def\vacs{V$_{\rm 606}$}
\def\iacs{i$_{\rm 776}$}
\def\zacs{z$_{\rm 850LP}$}

\newcommand{\oh}{\ifmmode 12 + \log({\rm O/H}) \else$12 + \log({\rm
O/H})$\fi}
\def\Oii{[O~{\sc ii}] $\lambda$3727}
\def\Oiii{[O~{\sc iii}] $\lambda\lambda$4959,5007}

\def\hyperz{{\em Hyperz}}
\def\flyf{\ifmmode f_{\rm Lyf} \else $f_{\rm Lyf}$\fi}
\def\pz{\ifmmode P(z) \else $P(z)$\fi}
\def\ki2{\ifmmode \chi^2 \else $\chi^2$\fi}
\def\zphot{\ifmmode z_{\rm phot} \else $z_{\rm phot}$\fi}

\newcommand{\xphot}{\ifmmode x_\gamma \else $v_\gamma$\fi}
\newcommand{\xobs}{\ifmmode x_{\rm obs} \else $x_{\rm obs}$\fi}
\newcommand{\xcmf}{\ifmmode x_{\rm CMF} \else $x_{\rm CMF}$\fi}
\newcommand{\vexp}{\ifmmode V_{\rm exp} \else $V_{\rm exp}$\fi}
\newcommand{\vmax}{\ifmmode V_{\rm max} \else $V_{\rm max}$\fi}
\newcommand{\nh}{\ifmmode N_{\rm HI} \else $N_{\rm HI}$\fi}
\newcommand{\dv}{\ifmmode \Delta v({\rm em-abs}) \else $\Delta v({\rm em}-{\rm abs})$\fi}

\def\fesc{\ifmmode f_{\rm esc} \else $f_{\rm esc}$\fi}

\def\frellya{\ifmmode f^{\rm rel}_{\rm{Ly}\alpha} \else $f^{\rm rel}_{\rm{Ly}\alpha}$\fi}

\def\whalpha{$W_{\rm{H}\alpha}$}

\def\ewha{$EW({\rm{H}\alpha})$}

\def\hii{H{\sc ii}}

\newcommand{\mstar}{\ifmmode M_\star \else $M_\star$\fi}
\newcommand{\muv}{\ifmmode M_{1500} \else $M_{1500}$\fi}
\newcommand{\luv}{\ifmmode L_{\rm UV} \else $L_{\rm UV}$\fi}
\newcommand{\lir}{\ifmmode L_{\rm IR} \else $L_{\rm IR}$\fi}
\newcommand{\lbol}{\ifmmode L_{\rm bol} \else $L_{\rm bol}$\fi}
\newcommand{\liruv}{\ifmmode L_{\rm IR+UV} \else $L_{\rm IR+UV}$\fi}
\newcommand{\liroveruv}{\ifmmode L_{\rm IR}/L_{\rm UV} \else $L_{\rm IR}/L_{\rm UV}$\fi}

\begin{document}
%
    \title{Properties of $z \sim$ 3--6 Lyman break galaxies. II.
     Testing star formation histories and the SFR--mass relation with ALMA and near-IR spectroscopy}
  \subtitle{}
  \author{D. Schaerer\inst{1,2}, S. de Barros\inst{1}, P. Sklias\inst{1}}
  \institute{
Observatoire de Gen\`eve, Universit\'e de Gen\`eve, 51 Ch. des Maillettes, 1290 Versoix, Switzerland
         \and
CNRS, IRAP, 14 Avenue E. Belin, 31400 Toulouse, France
 }

\authorrunning{}
\titlerunning{Testing star formation histories and the SFR--mass relation of Lyman break galaxies}

\date{Accepted for publication in A\&A}

{
\abstract{Understanding the relation between the star formation rate (SFR) and stellar mass (\mstar) of galaxies,
and the evolution of the specific star formation rate (sSFR=SFR/\mstar) .}
{We examine the dependence of derived physical parameters of distant Lyman break galaxies (LBGs) on the assumed star formation histories (SFHs), their
implications on the SFR--mass relation, and we propose observational tests to better constrain these quantities.}
{We use our SED-fitting tool including the effects of nebular emission to analyze a large sample of LBGs from redshift $z \sim 3$ to 6, assuming five different star formation histories, extending thereby our first analysis
of this sample (de Barros et al.\ 2012, paper I). In addition we predict the IR luminosities consistently with the SED fits.
}
{Compared to ``standard" SED fits assuming constant SFR and neglecting nebular lines, models assuming variable SFHs yield systematically
 lower stellar masses, higher extinction, higher SFR, higher IR luminosities, and a wider range of equivalent widths for optical emission lines. 
Exponentially declining and delayed SFHs yield basically identical results. Exponentially rising SFHs 
yield similar masses, but somewhat higher extinction than exponentially declining ones. 
We find significant deviations between the derived SFR and IR luminosity from the commonly used SFR(IR) or
SFR(IR+UV) calibration, due to differences in the SFHs and ages. 
 %
 Models with variable SFHs, favored statistically, yield generally a large scatter in the SFR--mass relation. We
 show the dependence of this scatter on assumptions of the SFH, the introduction of an age prior, and on the extinction law.
We show that the true scatter in the SFR--mass relation can be significantly larger than inferred using 
SFR(UV) and/or SFR(IR),  if the true star formation histories are variable and relatively young populations
are present.
We show that different SFHs, and hence differences in the derived SFR--mass relation and in the specific star formation rates,
can be tested/constrained observationally with future IR observations with ALMA.
Measurement of emission lines, such as \ha\ and \Oii, can also provide useful constraints
on the SED models, and hence test the predicted physical parameters. 
 }
{Our findings of a large scatter in the SFR--mass relation at high-z and an increase of the 
specific star formation rate above $z \ga 3$ (paper I) can be tested
observationally.
Consistent analysis of all the observables including the rest-frame UV to IR and emission lines are required
to establish more precisely true SFR values and the scatter in the SFR--mass relation.
}
 \keywords{ Galaxies: starburst -- Galaxies: ISM -- Galaxies: high-redshift --
Ultraviolet: galaxies -- Radiative transfer }

  \maketitle

\section{Introduction}
One of the important, recent results of multi-wavelength galaxy surveys is the finding of 
a well-defined relation between the star formation rate (SFR) and stellar mass of galaxies
-- now often called the main sequence of star forming galaxies --
and the evolution of this relation with redshift, at least from the nearby Universe up to redshift
$z \sim 2$ \citep{noeskeetal2007,daddietal2007,elbazetal2007,rodighieroetal2011}.
The observed redshift evolution corresponds to an increase of the typical specific star formation
rate (sSFR) by more than an order of magnitude from $z \sim 0$ to 2, indicating a higher star formation activity of galaxies
in the past.

Recent analysis of Lyman break galaxies (LBGs), combining often HST and Spitzer observations,
have tried to examine whether a SFR--mass relation is already in place at higher redshift, and
how the specific star formation rate of galaxies behaves at $z \sim$ 4--7
\citep[e.g.][]{stark09,schaerer&debarros2010,labbeetal2010,mclureetal2011,gonzalezetal2011}.
The first studies have found a relation similar to the $z \sim 2$ SFR--mass relation, which would 
indicate no evolution of the sSFR beyond redshift $\ga 2$ -- a plateau in sSFR --
\citep[e.g.][]{stark09,gonzalezetal2011}.
This result appears difficult to reconcile with most theoretical models, which successfully explain
the behavior of star formation properties from $z \sim$ 0 to 2, but predict a continuing
rise of the specific star formation rate towards higher redshift \citep{boucheetal2010,duttonetal2010,weinmannetal2011,daveetal2011},
although others propose alternatives \citep[e.g.]{krumholz&dekel2012}.
More recent work, including different spectral energy distribution (SED) models, or allowing for a revision of 
dust attenuation have shown that the observed sSFR at $z \sim$ 5--7 could be higher than
previously thought \citep{schaerer&debarros2010,debarros2011,bouwens2011_beta,yabeetal2009,dBSS12}.

If all galaxies at $z \ga 2$ obey a mass--SFR relation with star formation rate
increasing with galaxy mass
and its normalization remains basically unchanged with redshift, it is evident that their star formation rate must increases with time.
The apparent, small scatter in the SFR--mass relation, and the value of the exponent $\alpha$ 
close to unity in this relation SFR $\propto \mstar^\alpha$  found by various studies 
\citep[e.g.][]{daddietal2007,elbazetal2007, gonzalezetal2011} has led several authors to conclude that high redshift galaxies
must have gone through a phase of quasi-exponential growth \citep[e.g.][]{renzini2009,marastonetal2010}.
Earlier studies have independently advocated rising star formation histories (SFHs) for 
high redshift galaxies from hydrodynamic simulations and semi-analytical models
\citep{finlatoretal2007,finlatoretal2010,finlatoretal2011}.
Arguments in favor of rising SFHs, at least on average, have been put forward by
\citet{finkelsteinetal2010,papovichetal2011} to explain the evolution of the UV luminosity function.
If representative of the star formation history of individual galaxies, such star formation
histories would indeed by quite different from exponentially declining or constant SFHs,
most frequently assumed in  the literature to model/analyze the observed SED of distant galaxies. 

Assumptions on the star formation history affect the physical parameters of galaxies derived from 
SED fits. This is therefore generally treated as a free parameter. See e.g.\
\citet{reddy2010,marastonetal2010,wuytsetal2011,reddy2012}
for examples on $z \ga 2$ star forming galaxies.
Determinining star formation histories and their associated timescales is also important, 
as this may provide constraints on different modes of star formation and on feedback processes at high redshift,
recognized as key features for our understanding of galaxy evolution.
For example, star formation may proceed on different timescales if regulated by 
cold-accretion or by star formation feedback or triggered by mergers 
\citep[cf.][]{khochfar&silk2011,wyithe&loeb2011}.
These points already illustrate the interest and importance of determining
the star formation histories and typical timescales of star formation at high $z$.
We here wish to discuss and reexamine these issues on the basis of an analysis
of a large sample of LBGs covering redshifts from $\sim 3$ to 6, and to present 
observational tests which should be able to distinguish different star formation histories.

 One of main arguments often invoked to argue for rising star formation histories
 is the small scatter of the SFR--mass relation.  However, it should be recognized that
 generally the SFR is derived from observables -- the UV or IR luminosity -- which depend
 on relatively long ($\ga$ 100 Myr) timescales, and which are assumed to be at
 an equilibrium value, which is only reached after this timescale and for constant SFR.
 With these assumptions entering e.g.\ the commonly used SFR(UV) or SFR(IR) calibrations
 or \citet{kennicutt1998}, it is natural that the ``observational" scatter is smaller than the true scatter in 
 the current SFR, if typical ages are less than 100 Myr and/or the timescale shorter
 than this.  Especially at high redshift, where timescales are shorter 
 \citep[e.g.\ the dynamical timescale decreases with $(1+z)^{-3/2}$, cf.][]{wyithe&loeb2011}
one should therefore carefully
(re)examine the SFR--mass relation and its ``tightness" using consistent diagnostics, 
and with all the available observational constraints including on age, star formation
timescales and histories. 
In any case, the conclusion that rising SFHs are favored is {\em a priori} inconsistent with the
assumptions on the SFH for the determination of the SFR--mass relation and its scatter,
where most studies simply assume a constant star formation rate. 
Although this has been re-examined for $z\sim 2$ samples
\citep{marastonetal2010,wuytsetal2011,reddy2012} this obviously calls for a revision
at higher redshifts.

 Most SED studies of LBGs at $z \ga 3$ have assumed constant star formation
 rates, or exponentially declining SFHs to determine the physical parameters of these
 galaxies 
 \citep[e.g.][]{egamietal2005,SP05,eylesetal2005,vermaetal2007,yabeetal2009,stark09,leeetal2010,schaerer&debarros2010,gonzalezetal2011},
some of them imposing no dust attenuation, motivated by the blue UV slopes observed  at high redshift. 
 Notable examples are the work of \citet{finlatoretal2007} who analyzed 6 galaxies at $z>5.5$ with different SFHs,
 including rising ones taken from they hydro-dynamic simulations. The physical parameters they derive 
 are consistent with those using simple parametrised histories, albeit with reduced uncertainties.
 Their study also shows that SED fits with rising star formation histories yield
 a higher attenuation than inferred  assuming constant SFR, a result not yet appreciated enough,
 which we also find in \cite{dBSS12} and in this paper.
 Most recently, other groups have also analyzed high-$z$ samples with rising SFHs 
 \citep[e.g.][]{Curtis-Lake2012,Gonzalez2012}. However, dust extinction is, for example, not treated
 consistently with the star formation history in the approach of \cite{Gonzalez2012}.
 
Another drawback of most earlier studies is that the contribution of nebular emission (most emission lines)
is not taken into account, an effect which can significantly alter the ages, masses, and
other physical parameters derived from SED fits, as demonstrated by \cite{schaerer&debarros2009,
schaerer&debarros2010}. Indeed, there is now clear evidence for the presence of nebular lines
affecting the broad-band photometry of high-$z$ star-forming galaxies (Lyman alpha emitters
and LBGs in particular), as discussed e.g.\ in \citet{schaerer&debarros2011}.
The best demonstration comes from LBGs with spectroscopic redshifts
between 3.8 and 5, among which a large fraction shows a clear excess in the 3.6 \micron\ filter
with respect to neighboring filters (K and 4.5 \micron), as shown by \cite{shimetal2011}. 
At these redshifts \ha\ is located in the 3.6 \micron\ filter, whereas very few lines are expected
at 4.5 \micron, showing that the observed 3.6 \micron\ excess is naturally explained by strong
\ha\ emission.

Given these limitations of published SED studies and the interesting results obtained 
from our work on a small sample of $z \sim$ 6--8 LBGs \citep{schaerer&debarros2010},
we have recently undertaken an extensive study of the physical parameters of 
a large sample of $\sim 1400$ LBGs at $z \sim$ 3--6, using our state-of-the art photometric redshift 
and SED fitting model including nebular emission, and considering a range of different
star formation histories.
Among the numerous detailed results obtained in this work \citep[][hereafter dBSS12]{dBSS12}
we find in particular from the preferred models:
1) a large scatter in the SFR--mass relation for the preferred, variable star formation histories,
2) higher dust attenuation than obtained from models assuming SFR=const and from standard
methods using the UV slope, and
3) a higher sSFR than commonly obtained, and a rising sSFR with redshift.
Our models therefore reconcile the observationally-inferred specific star formation rate 
with predictions from cosmological models predicting a continuous rise of the sSFR with redshift 
\citep{boucheetal2010,duttonetal2010,weinmannetal2011,daveetal2011,krumholz&dekel2012}.
The results from dBSS12 have also other important implications, e.g.\ on the typical star formation
timescales at high redshift, and on the cosmic star formation rate density.
 
 Given the importance of these findings, it is of interest to carry out additional independent 
 tests of the models and to provide further constrains on the ages, extinction, and star formation histories
 of high redshift galaxies. In this paper we present predictions allowing such tests.
 Furthermore, we extend the study of dBSS12 by exploring other star formation histories, not
 considered in our previous paper. Indeed, while dBSS12 adopted the fixed, rising star history of 
\citet{finlatoretal2011} obtained from the hydrodynamic simulations, we here explore exponentially
 rising SFHs and so-called ``delayed" histories with variable timescales. 
 These histories were previously applied to the SED fits of galaxies at lower redshift,
  e.g.\ at $z \sim 2$ by \citet{marastonetal2010,wuytsetal2011,reddy2012}
  and found to be preferred over other simple star formation histories.
  
 First, we examine the effect of the different SFHs on the derived physical parameters of 
 LBGs. We then present the implications these different model assumptions have on the 
 SFR--mass relation. After that we present consistent predictions for the infrared luminosity
 of these galaxies, based on their SED fits and the assumption of energy conservation, 
 abandoning the standard SFR--\lir\ conversions. We show in particular
 that such a consistent prediction yields a smaller scatter in the observables (\lir) than in 
 the current SFR. The true star formation rate of LBGs at $z \ga 3$ could thus very well
 show a large scatter around a ``main sequence" even if the UV and/or IR luminosities
 show a small scatter. Our models also show that star formation histories can, to some 
 extend, be distinguished from measurement of IR luminosities, which will become
 possible with ALMA, since different amounts of UV attenuation are expected for different
 SFHs. We also present the predicted strength of some selected emission lines,
 which can be used to test our models and the different star formation histories.

Our paper is structured as follows. The observational data and the method used for SED modelling
are described in Sect.\ \ref{s_models}. The dependence of the physical parameters on the assumed star formation 
histories is discussed in Sect.\ \ref{s_phys}. Our general predictions for the IR emission are presented in Sect.\ \ref{s_ir}.
Specific predictions for $z \sim 4-6$ LBGs and ALMA are given in Sect.\ \ref{s_alma}.
The predicted strengths of optical emission lines are shown in Sect.\ \ref{s_lines}.
We discuss our results in Sect.\ \ref{s_discuss}, and Sect.\ \ref{s_conclude} summarises our main conclusions.
We adopt a $\Lambda$-CDM cosmological model with $H_{0}$=70 km s$^{-1}$ Mpc$^{-1}$, 
$\Omega_{m}$=0.3 and $\Omega_{\Lambda}$=0.7. 

\section{Observational data and SED modelling}
\label{s_models}

\citet[][hereafter dBSS12]{dBSS12} have analysed a large sample of $z \sim$ 3--6 
dropout-selected galaxies in depth using an up-to-date photometric redshift and SED-fitting tool,  that treats the 
effects of nebular emission on the SEDs of galaxies. 
In their homogeneous analysis they determine the main physical properties, such
as the star formation rate (SFR), stellar mass, age, and reddening.
They assess carefully their uncertainties, and discuss the evolution of these properties with redshift.
We here extend these simulations to include other star formation histories and we examine
the predicted IR luminosities and  equivalent widths of selected emission lines from these 
galaxies.

\subsection{Photometric data and sample selection}
We have used the GOODS-MUSIC catalogue of \citet{santini09}, which
provide photometry in the U, \bacs, \vacs, \iacs, \zacs, J, H, and K bands
mostly from the VLT and HST, and the 3.6, 4.5, 5.8, and 8.0 \micron\ bands
from the IRAC camera onboard {\em Spitzer}.
Using standard criteria as in \cite{noninoetal2009}  and \citet{stark09} we  then selected U, B, V, and i-drop 
galaxies. To reduce the contamination rate (typically $\sim$ 10--20 \%) we
only retained the objects whose median photometric redshifts agree 
with the targetted redshift range. This leaves us with a sample of 389, 705, 199, and 60 
galaxies with median photometric redshifts of $\zphot= 3.3$, 3.9, 4.9, and 6.0.
Typicallly, the completeness limit of our sample is $\muv \approx$ --19 
(corresponding to $\luv \approx 8 \times 10^9$ \lsun) at $z \sim 4$ and
up to $\muv \approx$ --19.8 at $z \sim 6$.  It is discussed in \citet{stark09}.
See dBSS12 for more details.

\subsection{SED models}
Our SED-fitting tool, described in \cite{schaerer&debarros2009} and \cite{schaerer&debarros2010},
is based on a version of the \hyperz\ photometric redshift code of \cite{bolzonellaetal2000}, 
modified to take nebular emission into account.
In dBSS12
we considered a large set of spectral templates based on
the GALAXEV synthesis models of \cite{bruzual&charlot2003} and covering
different metallicities and a wide range of star formation
histories. A Salpeter IMF from 0.1 to 100 \msun\ was adopted. Our results can
easily be rescaled to other IMFs with the same power law slope at high masses.
Nebular emission from continuum processes and numerous emission lines were added to the
spectra predicted from the GALAXEV models as described in
\cite{schaerer&debarros2009}, proportionally to the Lyman continuum
photon production. Following the results of dBSS12 we set the \lya\ line
to zero for $z\sim$ 3--5, and we assume the case B value for the i-drop sample.
A more detailed treatment of \lya, following e.g.\ \cite{schaereretal2011},
is not necessary and clearly beyond the scope of this paper, as the main physical parameters
show generally only a small dependence on the treatment of \lya.
The intergalactic medium (IGM) was treated with the prescription of \citet{Madau95}.

The free parameters of our SED fits are:
redshift $z$,  metallicity $Z$ (of stars and gas),  the age $t_\star$ defined since the onset of star-formation, 
the star formation history,
and attenuation $A_V$ described by the Calzetti law \citep{calzettietal2000}.
Stellar mass and the current star formation rate are derived from the absolute scaling of the SEDs.
For comparison, a subsample of the data was also modeled using the SMC extinction law
(cf.\ Sect.\ \ref{s_discuss}).

Since one of the main objectives of this paper is to explore a variety of star formation  histories (SFHs),
we have adopted five different histories, which are summarized in Table \ref{t_sfh}, and illustrated in Fig.\ \ref{fig_sfh}.
Note that for exponentially rising SFHs the SFR is set to zero, after a growth by more than 20 decades.
The bolometric luminosity output per unit SFR, respectively its inverse SFR/$L_{\rm bol}$, and the mass-to-light ratio in the $V$-band 
corresponding to these simple star formation histories are shown in Figs.\ \ref{fig_sfr_mbol}
and \ref{fig_masslight}. The behavior of these quantities are fundamental to explain the main
differences in star formation rate and stellar mass obtained from different models, discussed in
Sect.\ \ref{s_phys}.

The first three cases, constant star formation, exponentially declining SF, and the average rising SFH
predicted by the hydrodynamical models of \cite{finlatoretal2011} have already been used by
\citet{dBSS12}. Two additional SFHs, exponentially rising, and so-called ``delayed" SF 
showing a initial increase and subsequent decrease of the star formation,
are also considered here. 
The way these assumed SFHs modify the derived physical parameters of LBGs and how they
could be distinguished is discussed below.
 
\begin{figure}[htb]
\centering
\includegraphics[width=8.8cm]{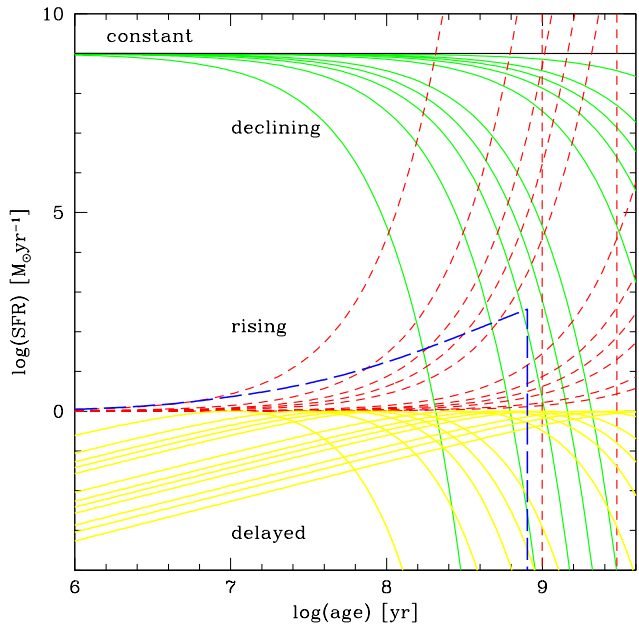}
\caption{Illustration of the star formation histories adopted in this paper. Shown is the SFR as a function of age.
Within each model set A--E the curves are normalized to an arbitrary value. The colors used for the different
SFHs are listed in Table \ref{t_sfh}.}
\label{fig_sfh}
\end{figure}

\begin{figure}[htb]
\centering
\includegraphics[width=8.8cm]{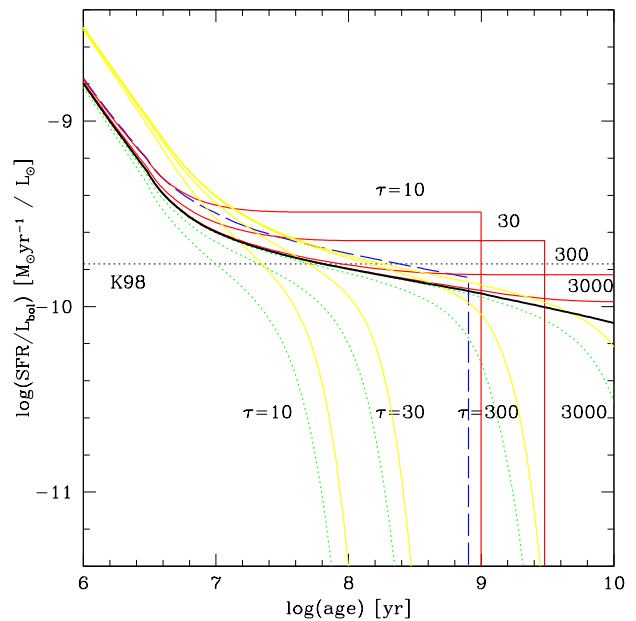}
\caption{Predicted ratio SFR/$L_{\rm bol}$ as function of time for the some of the models with exponentially decreasing, delayed,
and exponentially rising SFHs (with timescales $\tau=$ 10, 30, 300, 3000 Myr indicated), compared to the case of constant SFR. 
The plots are shown here for BC2003 models with solar metallicity. Same color codes as in Fig.\ \protect\ref{fig_sfh}.
The standard SFR(IR) conversion factor from \protect\cite{kennicutt1998} is shown by the black dotted line.}
\label{fig_sfr_mbol}
\end{figure}

\begin{figure}[htb]
\centering
\includegraphics[width=8.8cm]{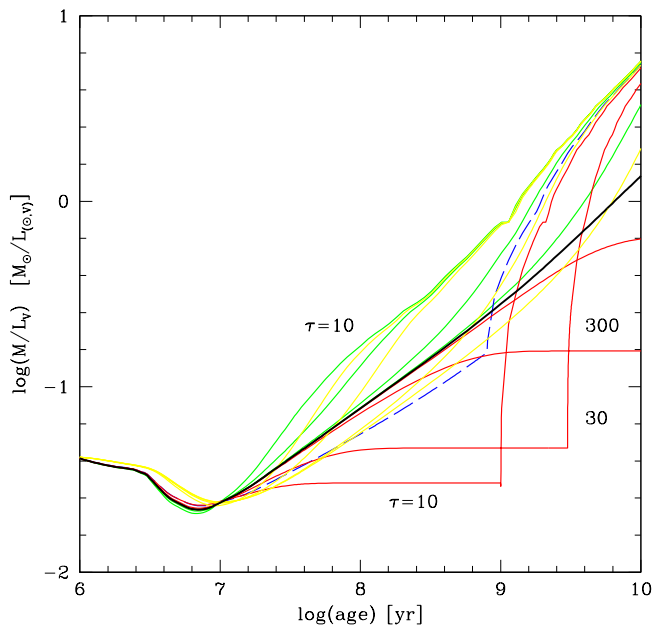}
\caption{Predicted mass-to-light ratio in the V-band, $\mstar/L_V$, for the same models as shown in Fig.\ \protect\ref{fig_sfr_mbol}.
Same color codes as in Fig.\ \protect\ref{fig_sfh}.}
\label{fig_masslight}
\end{figure}

\begin{table}[htdp]
\caption{Star formation histories modeled in this work. 
All describe the evolution of the SFR with time (age $t$, defined as the time since the onset of star formation).
Three of them have one additional free parameter, the SF timescale $\tau$. For model B a minimum age of 
50 Myr is assumed, for comparison with other studies.
Models A, B, and C are the same as in dBSS12.
The color code used in the figures is indicated in the last column.}
\begin{center}
\begin{tabular}{lllllllll}
ID & SF history & functional form & parameters & color \\
\hline
A=DEC & exp.\ declining & SFR$\propto \exp^{-t/\tau}$ & $t$, $\tau$ & green  \\
B=REF & constant & SFR=const                                 &$t \ge 50$ Myr& black \\
C=RIS & rising & Finlator+2011                      &$t$  & blue \\
D  & exp.\ rising & SFR$\propto \exp^{+t/\tau}$    & $t$, $\tau$ & red \\
E  & delayed & SFR$\propto t \exp^{-t/\tau}$     & $t$, $\tau$ & yellow\\
\end{tabular}
\end{center}
\label{t_sfh}
\end{table}

In practice we compute for each model set (i.e.\ star formation history) SED fits for all combinations of
$z \in $ [0,10] in steps of 0.1 
$Z=$ (0.02=\zsun, 0.004, 0.001),
$\tau=$ (0.01, 0.03, 0.05, 0.07, 0.1, 0.3, 0.5, 0.7, 1., 3., $\infty$) Gyr for model A\footnote{For exponentially 
rising SFHs we explore $\tau=$ (0.01, 0.03, 0.05, 0.07, 0.1, 0.3, 0.5, 0.7, 1., 2., 3.) Gyr,
for delayed histories $\tau=$ (0.01, 0.03, 0.05, 0.07, 0.1, 0.3, 0.5, 0.7, 1., 2., 3., 4., 5.) Gyr.},
51 age steps from 0 to the age of the Universe \citep[see][]{bolzonellaetal2000},
$A_V \in $ [0,2] mag in steps of 0.05.
Minimisation over the entire parameter space yields the best-fit parameters and SED, along with
other properties such as the stellar mass and star formation rate (SFR) and UV magnitude.
Other physical parameters, briefly described below, are determined from the best-fit SED.
To determine the uncertainties of the physical parameters, we used Monte-Carlo simulations
to generate 1000 realizations of each object perturbing the photometry in all filters according to 
the tabulated errors. Each of these 1000 realizations of each galaxy was fit. From the best-fit
results for each realization we derived the probability 
distribution function for each parameter/quantity, either for each individual object or for (sub)samples.

Simplifying assumptions made for our SED fits and possible caveats are discussed 
in Sect.\ \ref{s_caveats}.
 
\subsection{Predicted physical parameters and observables}
In addition to the fit parameters of our models, redshift $z$, metallicity $Z$, age $t$, SF timescale $\tau$,
stellar mass \mstar, and current star formation rate SFR, our models also allow us to determine other parameters and observables of interest. 
We briefly describe them here, for sake of clarity.
Note that for all the physical parameters and observables we determine the detailed 
probability distribution function (pdf) from the full MC simulations. From the 1D pdf we 
derive in particular the median value and the 68\% confidence interval, which we use
to illustrate the uncertainties on these parameters. For most of the paper we use
for simplicity the median value of the physical parameter of interest.

\subsubsection{Emission lines}
From the numerous emission lines included in our spectral templates we select some lines, primarily \lya, \ha, \hb, \Oii, \Oiii, 
for which we save the  emission line fluxes and equivalent widths predicted by the photometric SED fits. These lines can 
subsequently be used for comparison with spectroscopic observations (existing or future).

\subsubsection{UV and IR emission}
For straightforward comparison with observations we use the absolute UV magnitude \muv\
defined at 1500 \AA, and derived from the SED using a square filter of 300 \AA\ width centered on 
this wavelength.
To define an emergent (observed) UV luminosity \luv\ we follow common practice by computing
$\luv = \lambda F_\lambda$, where $\lambda=1800$ \AA, and $F_\lambda$ is the average flux
between 1400 and 2200 \AA.

Assuming energy conservation, i.e.\ that the attenuated/extinct stellar light is reemitted by dust in the IR, 
we can predict the amount of IR radiation expected from each galaxy. The corresponding IR luminosity \lir\
is simply computed from the difference between the intrinsic, unattenuated SED 
given by the spectral template, and the best-fit SED with the parameters. 
In practice  \lir\ is calculated from the attenuated energy integrated from 913 \AA\ to 3 \micron\
following the same procedure as SED models carrying out complete from the UV-visible-IR fits 
\citep[cf.][]{dacunha2008,noll2009}. This quantity also includes the nebular emission (lines and continuum) if present in the
SED, consistently with our assumption of identical attenuation for both stars and gas.
We have checked that the integral over the Balmer continuum and up to 3 \micron\ yield 
basically identical results.

The method used to compute the IR luminosity does not make any explicit assumption 
on the origin of the stellar sources contributing to it and on the timescale over which the IR
emission is produced. All the energy emitted between the UV and 3 \micron\ and attenuated by dust 
according to the adopted extinction/attenuation law contributes to the IR emission, irrespective of 
stellar type, age etc. In practice, however, the timescale of the IR emission is not very different
than that of UV emission for the cases modeled here, since the bulk of the emitted energy 
originates from the UV, where the attenuation is also strongest.
In models assuming a specific dust geometry, time-dependent attenuation, multiple populations
or alike \citep[e.g. the models of ][]{dacunha2008}, the timescale of UV and IR emission could be different. 
However, such models include additional free parameters, and have, to the best of our knowledge, 
not yet been used to fit  distant galaxies. For simplicity we here fit the SEDs with single stellar populations 
and a single extinction/attenuation law.

From the UV and IR luminosity just defined, we also compute \liruv, the emergent luminosity
in the UV + IR domain, which is a good proxy of the total bolometric luminosity and 
to some extent also of the SFR (but cf.\ below).
Similarly, we will use the ratio of the IR to UV luminosity, \liroveruv, a quantity known
to be a good measure of dust attenuation \citep[cf.][]{Meurer99}.

\subsubsection{Predicted IR, sub-mm, and mm fluxes}
Given the IR luminosity \lir\ predicted consistently from the amount of attenuation by dust,
it is straightforward to predict the flux at various IR wavelengths, provided a corresponding
spectral template for dust emission is adopted. We simply assumed standard modified
black-body spectra described by $F_\nu \propto \nu^\beta B_\nu(T_d)$, 
with a fixed value of $\beta=2$ and three different dust temperatures $T_d=(25,35,45)$ K,
which span the range of observed dust temperatures in nearby and distant galaxies.
The modified black-body is scaled to \lir\ computed from 8 to 1000 \micron\ (rest frame).
From this, we determine the predicted IR fluxes in bands corresponding to
observations with Herschel (PACS and SPIRE bands), several sub-mm bands (350, 450, 850, 870 \micron),
1.2 and 2 mm bands, and ALMA bands 1-10 (corresponding to $\sim$ 8.1 to 0.35 mm). 
Empirical or semi-empirical spectral templates covering the IR domain could
also be used to convert the predicted IR luminosities to observed fluxes.
However, since it is not known how well the empirical templates based on low redshift
and/or high IR luminosity galaxies describe typical LBGs at $z>3$, and to use a simple
prescription, we have used modified black-body spectra. In any case, our conclusions
rely on the total IR luminosity, which is template independent.
Of course, the predicted \lir\ depends indirectly on the  stellar templates
and attenuation law assumed for the SED fits.

\section{Dependence of the physical parameters on the assumed star formation histories}
\label{s_phys}

To illustrate and discuss the dependence of the physical parameters on the assumed star formation histories
we here choose the largest sample comprising 705 B-drop galaxies. 
The general behavior discussed here is also found for most of the other samples at $z \sim 3$, 5, and 6.
Some results for other redshifts are discussed in Sect.\ \ref{s_alma}.
First we show briefly how the SFHs affect the derived physical parameters. Then we compare the
fit quality of the models, and discuss implications on the SFR--mass relation.

\subsection{Comparison of physical parameters}

It is well known that the physical parameters determined from broad-band SED fits depend on assumptions
such as the star formation history. In dBSS12 we have shown how quantities such as the stellar mass,
age, dust reddening,  and SFR depend on this assumption and how they are modified with the
inclusion of nebular emission.
Here we briefly illustrate the dependence of the main physical parameters on SFH, including in particular
the new histories explored in this paper (exponentially rising and delayed SFH; cf.\ Table \ref{t_sfh}).
The effects of nebular emission on the SED fits and the resulting physical parameters have been
discussed in detail in paper I (dBSS12). In particular, we have shown in paper I that nebular
emission affects approximately two thirds of the galaxies at all the redshifts considered ($z \sim$ 3--6),
quite irrespectively of their UV magnitude and galaxy mass.
For more details see paper I.

\begin{figure}[htb]
\centering
\includegraphics[width=8.8cm]{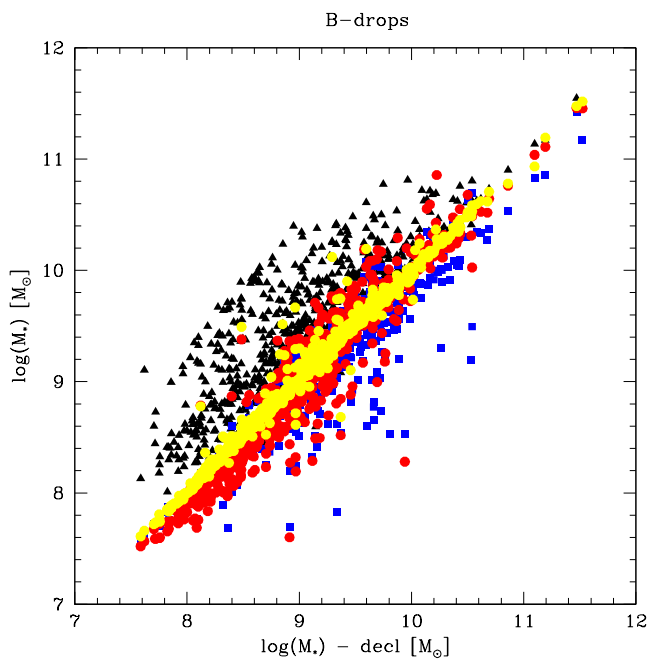}
\caption{Comparison of the median stellar masses of our $z \sim 4$ sample with 705 galaxies derived from SED fits
assuming the star formation histories listed in Table \ref{t_sfh}. The masses are compared to those derived
from model A, plotted on the x-axis. Black triangles show model B with constant SFR, blue and red symbols the rising
histories (model C and D respectively), yellow circles show the delayed SFHs (model E). For each model the same 
number of galaxies are plotted. Invisible data points are hidden close to the one-to-one relation.}
\label{fig_mstar}
\end{figure}
\begin{figure}[htb]
\centering
\includegraphics[width=8.8cm]{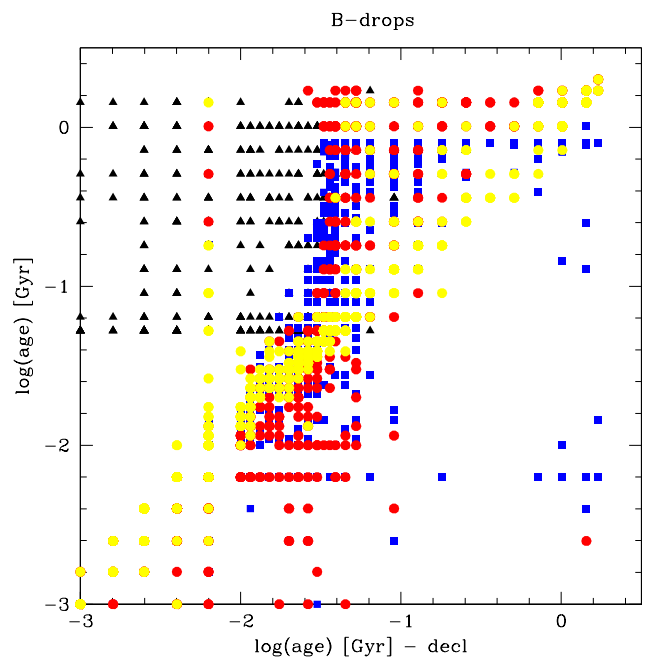}
\caption{Same as Fig.\ \protect\ref{fig_mstar} for the stellar age, defined since the onset of star formation.
For each model the same number of galaxies are plotted. Invisible data points may be hidden, especially under the yellow symbols of 
model E plotted last, due to discrete age values.}
\label{fig_age}
\end{figure}
\begin{figure}[htb]
\centering
\includegraphics[width=8.8cm]{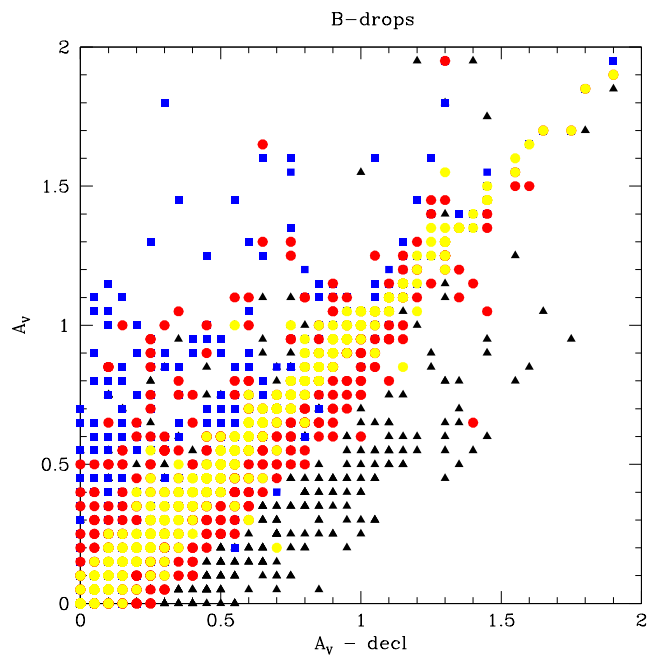}
\caption{Same as Fig.\ \protect\ref{fig_mstar} for the attenuation \av.
For each model the same number of galaxies are plotted. Invisible data points may be hidden, especially under the yellow symbols of 
model E plotted last, due to discrete age values.}
\label{fig_av}
\end{figure}
\begin{figure}[htb]
\centering
\includegraphics[width=8.8cm]{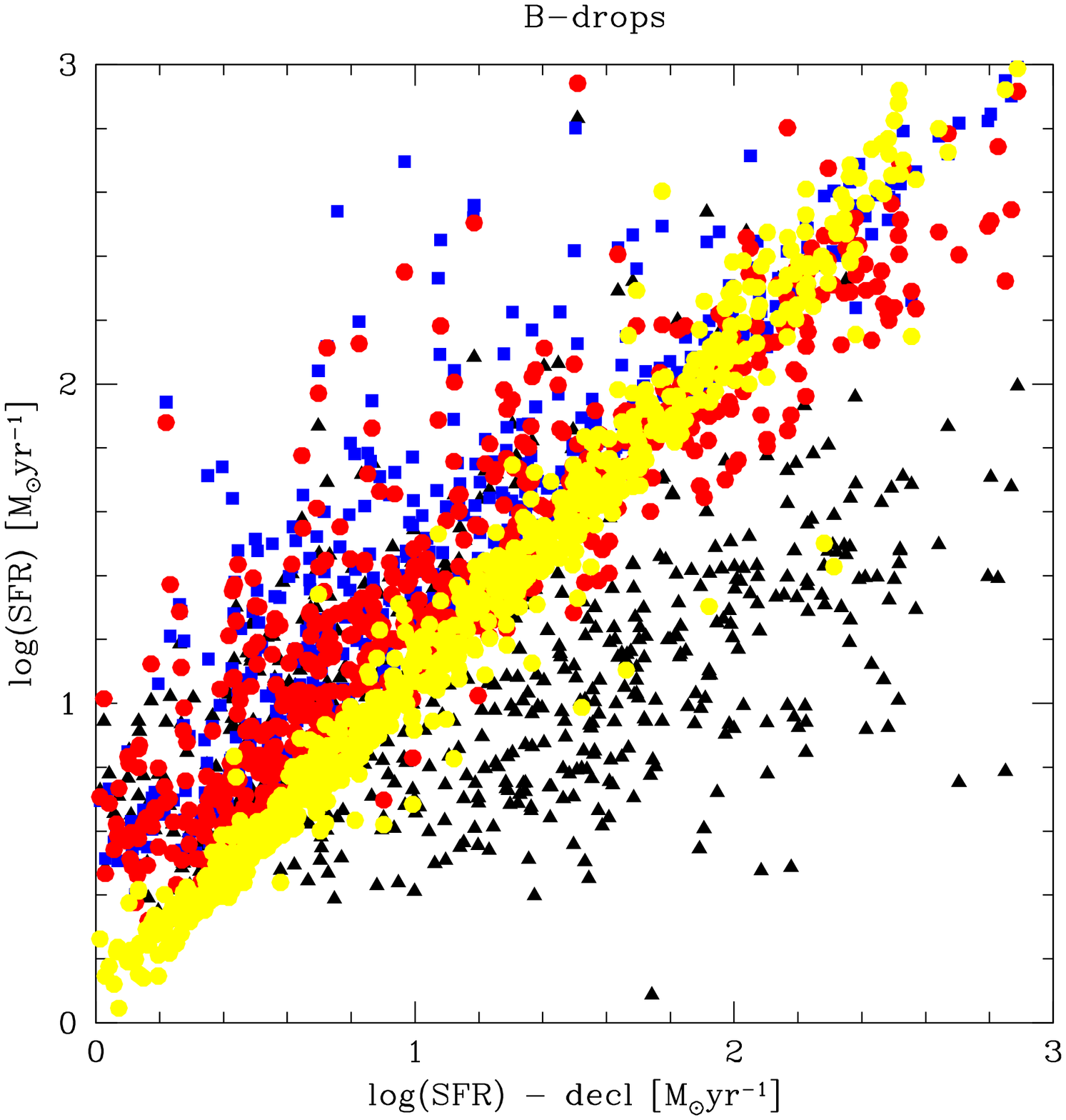}
\caption{Same as Fig.\ \protect\ref{fig_mstar} for the current star formation rate SFR.}
\label{fig_sfr}
\end{figure}

Figures \ref{fig_mstar} to \ref{fig_sfr} show how the \mstar, age, \av, and SFR derived for different 
assumptions on SFH compare to models assuming exponentially declining star formation histories
(our model A) for our largest sample, the 705 B-drop galaxies. 
This model is taken as a reference, since it generally provides the best fits for the vast
majority of galaxies analyzed here (cf.\ below). For all models except for SFR=const (model B), 
we include  the effects of nebular emission. Model B, called REF in dBSS12, serves as a
``reference'' model to ease comparisons with other studies in the literature assuming constant
star formation, and neglecting nebular emission. 
When nebular emission is added to model B, this does not lead to very different parameters
(cf.\ dBSS12).

Figure \ref{fig_mstar} shows that most models yield similar stellar masses,
as already known from earlier studies \citep{yabeetal2009}. The largest differences are found with respect to
models assuming  constant star formation and a mimum age of 50 Myr (model B), which yield generally higher
stellar masses. On average we find masses lower by a factor $\sim 3$ at $z \sim 4$ for declining 
SFHs (model A), where half of this effect is due to inclusion of nebular emission \citep[cf.][]{dBSS12}.
The difference is mostly due to the average shift towards younger ages, which have a lower mass-to-light
ratio, as shown in Fig.\ \ref{fig_masslight}.
It is immediately apparent that the models with delayed star formation histories (model E) yields very 
similar results to model A. Exponentially rising SFHs with variable timescales (model D) 
and the rapidly rising history from \cite{finlatoretal2011} (model C) give also comparable
stellar masses, although with a larger scatter, when compared to model A. 
In short, all the models with variable SFHs and nebular emission considered here yield quite similar stellar masses,
which are on average lower than those obtained from standard models without nebular emission.

Age is notoriously known to be  sensitive to models assumptions \citep[e.g.][]{papovichetal2001,yabeetal2009}. 
Indeed, absolute ages predicted 
by different models can differ by orders of magnitude,  as shown in Fig.\ \ref{fig_age}.
Overall models with delayed SF (E) show again quite similar results to those with exponentially decreasing 
histories (A). The behavior of the rising SFHs (C and D) is quite comparable. By construction, model B yields
only stellar ages $\ge 50$ Myr\footnote{See Sect.\ \ref{s_agemin} for a discussion of the age limit and other
implications.}
Since recent populations dominate for exponentially rising star formation,
ages are more difficult to constrain. This is reflected in increased uncertainties, e.g.\ by wider confidence
intervals (not shown here), as we have verified from our models.
Whereas for exponentially declining and delayed SFHs most objects are best fit with relatively short and comparable timescales 
$\tau$, the opposite is true for fits with exponentially rising histories. In this case the probability distribution function
for $\tau$ clearly peaks at the maximum value allowed here ($\tau=3$ Gyr), which means that a relatively
slow growth is preferred for most galaxies. 
This seems quite natural given the finding that the rising SFH from
\cite{finlatoretal2011}  fits on average less well than declining models.

The attenuation predicted by the various models is shown in Fig.\ \ref{fig_av}.  As already discussed by
dBSS12, constant SF models show generally lower dust attenuation than the exponentially declining models,
whereas the rising SFH of \cite{finlatoretal2011} predicts a higher attenuation on average. Rising star formation
histories with variable timescales globally scatter around those assuming declining SF.
Again, delayed SFHs give results very similar to those with declining SFHs. 

The predicted median SFR of all galaxies are compared in Fig.\ \ref{fig_sfr}. As expected due to 
lower dust attenuation and older ages, the models with constant SFR show generally the lowest 
(current) star formation rates (cf.\ dBSS12). As before, models A and E are very similar.
Finally, assuming rising star formation histories generally implies a higher SFR, 
due to, first, the lower \luv\ output per unit SFR (cf.\ Fig.\ \ref{fig_sfr_mbol}),
and, second, to a higher average dust attenuation, needed to redden the UV-to-optical colors, which are
always dominated by the youngest stars.
Again, this result is discussed in dBSS12, and has already been pointed out earlier by
\cite{SP05} and shown in the analysis of  \citet{finlatoretal2007}.

For all models with no age constraint, fits at young age will generally push the derived
SFR upward. For this reason we discuss below (Sect.\ \ref{s_agemin}) how the SFR changes when an age prior 
is adopted. The age distributions of our samples found for different SFHs and at different redshifts
are extensively illustrated in paper I (dBSS12), which also discusses possible caveats about the age. 
Overall, our models show a trend of increasing median age with decreasing redshift (e.g.\ at fixed absolute UV magnitude).
At $z \sim 3$ e.g., the typical median ages are 30 (90) Myr for models with declining (rising) SFH
and including nebular emission, compared to $\sim$ 250 Myr for model B (constant SFR, no lines), 
comparable to others in the literature. Obtaining independent age constraints should also help
testing the various models, and narrowing down the SFR differences between models.

\begin{figure}[tb]
\centering
\includegraphics[width=8.8cm]{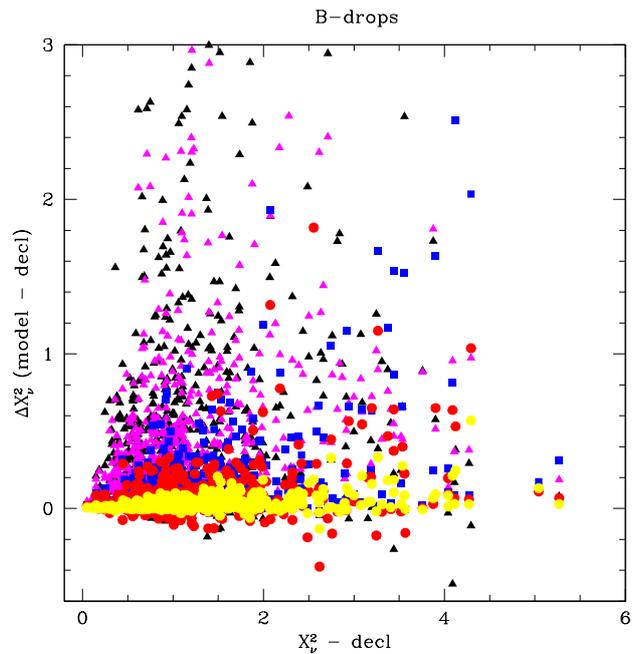}
\caption{Comparison of the median value of the reduced $\chi_\nu^2$ for each galaxy of the B-drop sample ($z \sim 4$)
computed for different star formation histories and plotted as a function of $\chi_\nu^2$ obtained from 
model A (declining star formation histories). Same symbols and colors as in previous figures and listed in 
Table \protect\ref{t_sfh}.} 
\label{fig_chi2}
\end{figure}

\subsection{Comparison of the fit quality}
Figure \ref{fig_chi2} shows the fit quality of the models expressed by the reduced  $\chi_\nu^2$ value
\footnote{Here $\chi_\nu^2$ is defined as $\chi^2/(N_{\rm filt}-1)$, where $N_{\rm filt}$ is the number of filters
for which data is available for the galaxy, including non-detections.} for the $z \sim 4$ sample. 
Overall, declining star formation histories (model A) give the best fit to most galaxies, and delayed SFHs very similar fits. 
Exponentially rising histories provide better fits for some objects, but a slightly higher $\chi^2$ on average.
Models B and C with constant SFR and rapidly rising star formation yield fits of lower quality for most
galaxies.
Overall the significance of the fit quality is relative small, as can be seen by the \ki2\ differences (Fig.\ \ref{fig_chi2}).
Since model B (constant SFR) is a subset of model A (exponentially declining), the difference between two models can e.g.\
be compared using the likelihood ratio test. The average (median) \ki2\ difference between the two models
corresponds to a probability of $\sim$ 25--30\% for model B being a better fit than model A.
Statistically speaking, based on the fitted part of the SED, it is therefore not possible to firmly
distinguish different SFHs. Other constraints, as the IR emission or emission line measurements,
will be needed, as discussed in the rest of the paper. See also dBSS12 for other arguments 
on different star formation histories.

As also done in other studies, we have examined the influence of another reddening law (cf.\ Sect.\  \ref{s_discuss}). 
Applying the SMC law of \cite{prevotetal1984,bouchetetal1985} we find that the majority of galaxies fit less well than 
with the Calzetti law, both the declining and rising star formation histories (model A, C, and D),
although the differences are generally relatively small (cf.\ Sect.\ \ref{s_discuss}).
Some studies have favored the SMC law for a fraction of LBGs e.g.\ for young objects
\citep[cf.][]{reddy2010,shimetal2011}, whereas usually the Calzetti law is thought to be a good description of the observations.

\subsection{Implications for the SFR--mass relation and influence of the minimum age}
\label{s_agemin}

Figure \ref{fig_mstar_sfr} shows the SFR--mass relation for the 705 B-drop galaxies
as derived from the SED fits assuming the five different star formation histories.
As already discussed in dBSS12, the models show overall a large scatter and a median
specific SFR (SFR/\mstar) above the mean relation derived by \citet{daddietal2007} at $z \sim 2$,
also shown here comparison with other studies and redshifts.
In part the scatter in this diagram is determined by the possible range of SFR/\mstar\ predicted
for stellar populations with different SFHs and different ages. Indeed, for a given $\lbol/L_V$,
probed observationally by $\luv/L_V$,
the ratio SFR/\mstar\ varies by $\sim$ 2 orders of magnitude between young ages and $t \la 1$ Gyr 
for constant SFR models, and more for declining or delayed star formation histories (Figs.\
\ref{fig_sfr_mbol} and \ref{fig_masslight}). Of course, since reddening also modifies $\luv/L_V$,
this adds further scatter.

As expected, the delayed SFH yields basically identical results to the declining histories.
Rising histories (model D) occupy a similar domain as the other variable SFHs (models A, C, E),
except for the {\em absence of galaxies below} the mean relation from  \citet{daddietal2007}. 
This is simply due to the fact that star formation cannot decrease in this model and for 
model B (constant SFR). Hence situations where the current SFR is low, but the UV flux is still 
present from somewhat older stars (e.g.\ B-type stars with lifetimes up to $\sim$ 100 Myr) 
are not possible for these SFHs (cf.\ Fig.\ \ref{fig_sfr_mbol}). 
Objects well below the SFR--mass relation require obviously decreasing star formation
histories, or some shut-off of star formation.

The {\em upper boundary} of the SFR--\mstar\ plot, parallel to the indicated $z \sim 2$ relation, is determined by 
the minimum age, and by dust attenuation. 
Indeed, young ages imply lower masses and a higher SFR from simple considerations of SFR/UV and 
mass/light ratios 
 \citep[cf.\ Figs.\ \ref{fig_sfr_mbol} and \ref{fig_masslight}, also][]{mclureetal2011}. 

Since age is kept free in all our models\footnote{The only age limitation is given by the age of the Universe.}
except for model B where we assume a minimum age of 50 Myr, and since some the median ages
obtained are relatively young (cf.\ dBSS12) it is interesting to explore how the main physical parameters and the SFR--\mstar\ relation
change when introducing a prior on the minimum age (or equivalently on the formation redshift) for all models.
The physical motivation for such a minimum age of 50 Myr is to avoid young ages, which 
may be unrealistic for the entire stellar populations of a galaxy, in particular in view of longer dynamical timescales
\citep[e.g.][]{reddy2010}.

The effect of fixing the minimum age to $t_{\rm min}=50$ Myr in the SED fits with exponentially declining and 
rising star formation histories (models A and D) on the SFR--mass relation is shown in Fig. \ref{fig_mstar_sfr_agemin}.
In short, for the declining SFHs the {\em upper boundary} becomes now identical to the one obtained
with SFR=const (model B), as expected, since declining SFHs include long timescales $\tau$ for which
the star formation history becomes undistinguishable from constant, and since $t_{\rm min}=50$ Myr
sets the maximum SFR/UV output. 
In contrast, for rising SFHs the upper boundary is found to be higher by a factor $\sim 3$ approximately.
Again, this simply due to the less efficient UV output per unit SFR (or higher SFR/UV ratio) of the youngest stars, 
which always dominate the SFR for such star formation histories (Fig.\ \ref{fig_sfr_mbol}).
The {\em lower boundary} of the SFR--mass relation remains of course unchanged by changes of the minimum age.
In conclusion, introducing a minimum age affects in different ways the SFR--\mstar\ relation, depending
on the SFH.  

Among the $z \sim 4$ sample, 50-70 \% have median ages less than 50 Myr for models with nebular
emission and for the different SFHs. As already mentioned, the uncertainties on ages are large. 
The median ages are also found to increase with decreasing redshift and with increasing mass, as shown in dBSS12.
On average, i.e.\ for the entire sample of 705 B-drop galaxies, the median age and mass increase by a factor $\sim 2$, 
whereas SFR and $A_V$ decrease by a factor $\sim 2$, when the prior $t_{\rm min}=50$ Myr is introduced.
For the rising histories (model D) the median age increases by a factor $\sim 5$, 
SFR decreases by 1.5, stellar mass increases by a factor 2. On the other hand the dust attenuation, both of
individual galaxies and on average, remain very similar, with interesting implications on the predicted IR luminosity
(see Sect.\ \ref{s_minage}).

To conclude this section we note that SED fits using models with variable star formation histories
imply a fairly large scatter in the SFR--mass relation for LBGs at $z \ge 3$, as shown here and in dBSS12.
However, it must be pointed out that the amount of scatter cannot directly be translated into observable
quantities such as UV and/or IR luminosity, which are commonly used as SFR indicators. 
Indeed, as we will show below (Sect. \ref{s_liruv_mstar}), a much smaller scatter is expected in terms of 
observed UV+IR luminosity. 

\begin{figure}[htb]
\centering
\includegraphics[width=8.8cm]{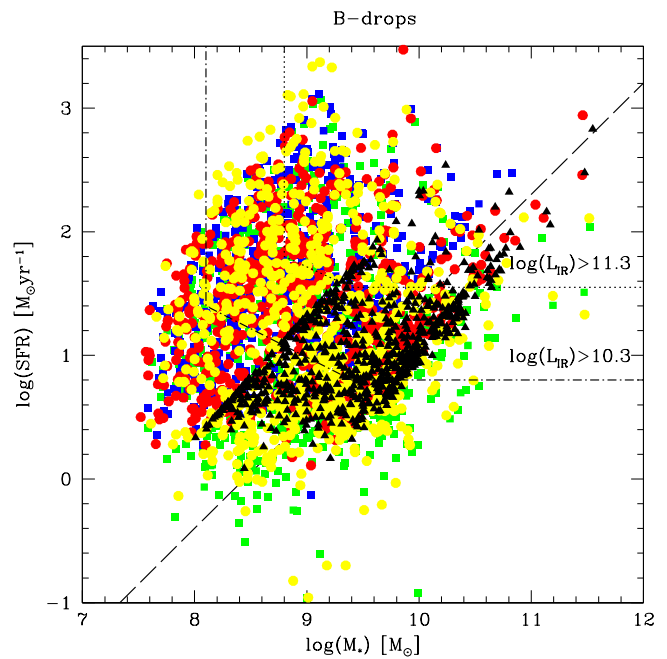}
\caption{Current SFR as a function of stellar mass of the $z \sim 4$ LBGs obtained for different SF histories.
Same symbols as Fig.\ \protect\ref{fig_mstar}. The dotted and dash-dotted lines delimit the area above which galaxies
with $\lir \protect\ga 2.\times 10^{11}$ \lsun\ (or equivalently a flux $>0.2$ mJy in ALMA band 7 
(approximately at band center, $\sim 950$ \micron)  for $T_d=35$ K)
or $>2.\times 10^{10}$ \lsun, respectively, are found.
The dashed line shows the SFR--mass relation at $z \sim 2$ for comparison.} 
\label{fig_mstar_sfr}
\end{figure}

\begin{figure}[htb]
\centering
\includegraphics[width=8.8cm]{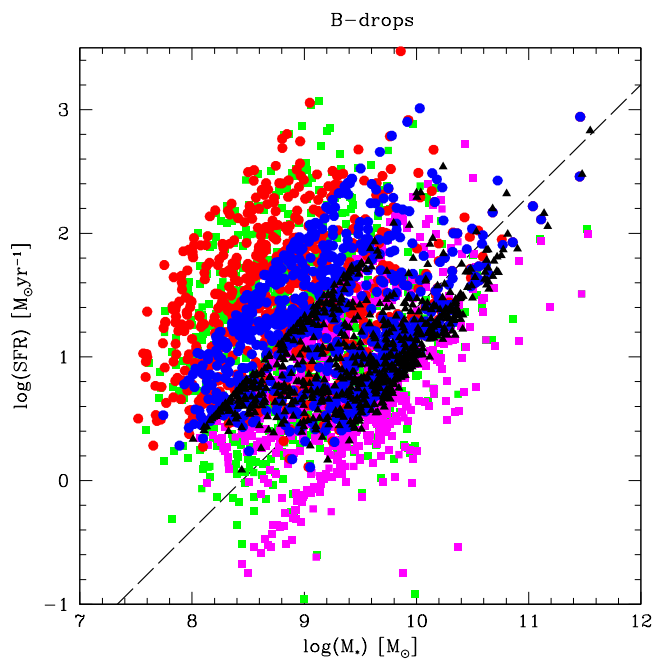}
\caption{Effect of adopting a minimum age prior on the SFR-mass relation.
Green (red) show the models with declining (rising) SFHs, as in Fig.\ \protect\ref{fig_mstar_sfr}.
Including the prior $t>50$ Myr yields the magenta (dark blue) points for the declining (rising) SFHs.
Note the decrease of the upper boundary of the SFR--\protect\mstar\ relation, but less so for rising SFH than for declining ones.
Black triangles stand for model B with SFR=const. 
The dashed line shows the SFR--mass relation at $z \sim 2$ for comparison.
}
\label{fig_mstar_sfr_agemin}
\end{figure}


\section{Predicted IR-mm emission}
\label{s_ir}
As shown in the previous section, the physical parameters inferred from SED fits depend in particular on
the assumed star formation histories and on possible age priors (assumptions of a minimum age).
We now present the IR properties predicted from the SED fits and discuss their dependence on 
these parameter. The effects of different extinction laws are addressed in Sect.\ \ref{s_discuss}.

\subsection{IR luminosity and constraints on UV attenuation by dust}

Figure \ref{fig_lbol} shows the infrared luminosity predicted by the different models for the
$z \sim 4$ sample, Fig.\ \ref{fig_iruv} the corresponding histogram of \liroveruv\footnote{Here
\luv\ is essentially model independent (except for the photometric redshift), since determined by the observed
UV magnitude.}.
Since some galaxies have a median $A_V=0$ and hence no UV light is
absorbed, their predicted median IR luminosity is $\lir=0$. Such galaxies are not shown on these
plots; their percentage ranges from 24\% for model B (with constant SFR) to 10\% for model C,
which has the highest median \lir.
Due to the adopted discretisation in $A_V$, the lowest, non-zero extinction $A_V=0.05$
corresponds to $\log(\lir/\luv) \approx -0.85$, as seen e.g.\ in Fig.\ \ref{fig_beta_iruv}.

As expected from the behavior of the UV attenuation discussed above and in dBSS12,  the models with constant SFR
predict generally significantly lower \lir\ than models with different star formation histories.
On average, the highest \lir\ is predicted for steeply rising SF of \citet{finlatoretal2011}, followed
by the exponentially rising model D, and model A and E with very similar average IR luminosities. 
Clearly, if \lir\ can be determined with sufficient accuracy from observations, it should be possible
to distinguish different models for individual galaxies, as well as statistically for large enough samples.

Indeed, the IR/UV ratio allows one to measure the UV attenuation 
as well known \citep[cf.][and references therein]{buatetal2005,igleasias-paramo2007}, and this independently 
of the assumed SF history and of the extinction law.
The corresponding relation between the UV attenuation factor $f_{\rm UV}$ and the IR/UV ratio is
shown in Fig.\ \ref{fig_att} for all galaxies of the $z \sim 4$ sample,
for all star formation histories and for the Calzetti attenuation and SMC extinction curve.
Here $f_{\rm UV}$ is defined by the ratio of the intrinsic (emitted) over the observed UV luminosity, i.e.\
\begin{equation}
\label{eq_fuv}
f_{\rm UV} = \luv^{\rm int} / \luv.
\end{equation}
The UV attenuation factor (defined here at 1800 \AA) can simply be described by 
\begin{equation}
\log(f_{\rm UV}) = 0.24 + 0.44 x + 0.16 x^2
\end{equation}
where $x=\log(\lir/\luv)$. For a specific attenuation law this can of course be translated into
quantities such as \av. For example, for the Calzetti law one has $A_V = 2.5 (R_V/k_\lambda) \log f_{\rm UV}= 1.08 \log f_{\rm UV}$.
%
Such a simple relation is quite naturally obtained, since the bulk of the 
energy reemitted in the IR after absorption by dust is also produced by the same stars
contributing the bulk of the UV emission. More complicated relations could be 
expected when e.g.\ older stellar populations such as AGB or RGB stars contribute
to dust emission, as appears to be the case in low redshift galaxies \citep{buatetal2005}.

In short, measurements of \lir/\luv\ of individual LBGs provide a direct measure of their
UV attenuation, which should allow us to distinguish certain star formation histories, 
and in particular to test the systematic shift in physical properties predicted between 
models assuming long star formation timescales (constant SFR) and variable (rising
or declining) star formation histories.
However, additional observational constraints, such as emission line strengths discussed later, 
may be necessary to distinguish between among variable SFHs, since both
rising and declining histories yield fairly similar predictions for many physical parameters
(cf.\ dBSS12 and above).

\begin{figure}[htb]
\centering
\includegraphics[width=8.8cm]{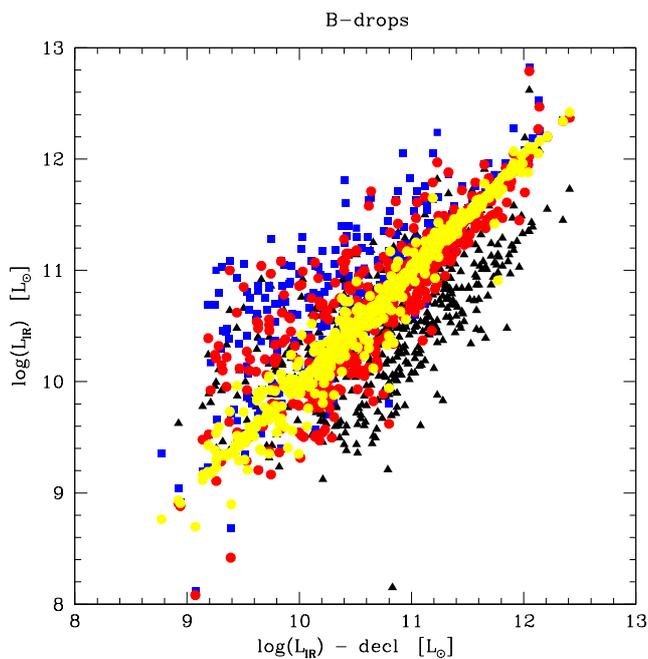}
\caption{Same as Fig.\ \protect\ref{fig_mstar} for the predicted IR luminosity.}
\label{fig_lbol}
\end{figure}

\begin{figure}[htb]
\centering
\includegraphics[width=8.8cm]{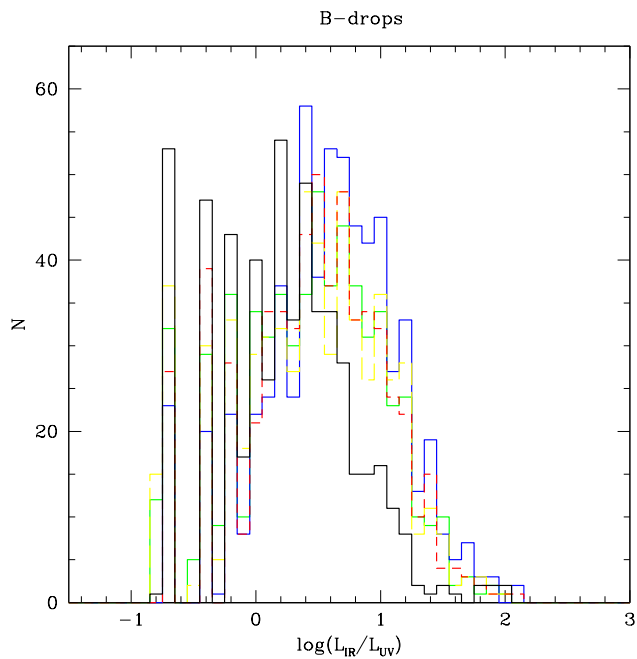}
\caption{Histogram of the predicted median IR/UV luminosity ratio for the $z \sim 4$ sample
from models assuming different star formation histories (Model A: green solid line;
model B: solid black; model C: solid blue; model D: red dashed; model E:  yellow dashed).} 
\label{fig_iruv}
\end{figure}

\begin{figure}[htb]
\centering
\includegraphics[width=8.8cm]{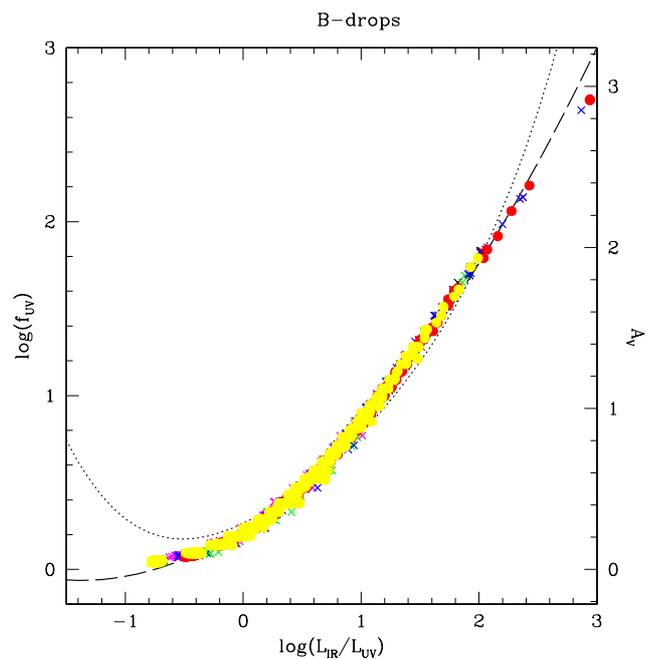}
\caption{UV attenuation factor $f_{\rm UV}$ as a function of the predicted ratio of the IR/UV luminosity.
A tight relation, fitted by Eq.\ \protect\ref{eq_fuv} and shown by the dashed line, is obtained independently of the assumed SF history and of the extinction law.
For comparison the relation determined by \citet{boquienetal2012}  is shown by the dotted line.
The corresponding attenuation $A_V$ assuming the Calzetti law is shown on the right scale.
} 
\label{fig_att}
\end{figure}

\subsection{IR and UV luminosity as tracers of the SFR in light of different star formation histories}
\label{s_iruv}

Often the IR luminosity or the sum of IR+UV luminosity are used as a measure of the 
star formation rate \citep[e.g.][]{kennicutt1998,burgarellaetal2005,reddy2010,reddy2012}.
The basic assumptions made 
for such conversions are that of constant SFR and an age of $\ga$ 50--100 Myr, necessary
to reach an equilibrium output for the UV and bolometric luminosities (see also Fig.\ \ref{fig_sfr_mbol}). 
The determination
of the SFR from \lir\ also assumes that the total (stellar) bolometric luminosity is
reemitted in the infrared domain, which requires significant dust attenuation. Obviously
the above assumptions cannot be valid in general, and in particular they are not 
compatible with the different star formation histories explored in this paper.

\begin{figure}[htb]
\centering
\includegraphics[width=8.8cm]{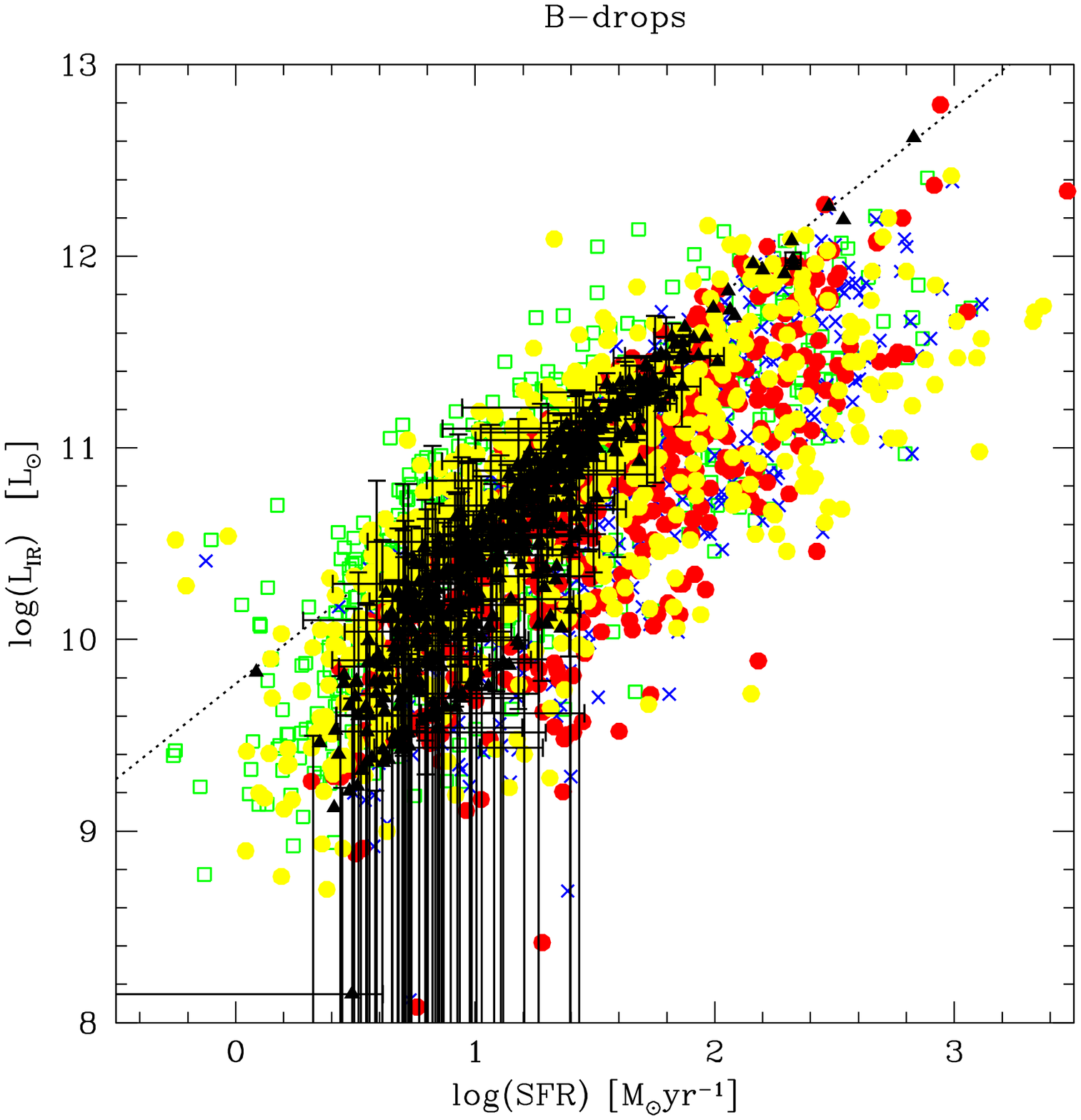}
\caption{Median IR luminosity \lir\ predicted for the B-drop sample as a function of the median SFR
obtained from the SED fits for the models with different star formation histories. 
Results from exponentially declining models (A) are plotted with green squares.
The remaining symbols and colors are the same as in previous figures.
For the constant SFR model the results, shown as black triangles, we also plot the 68\%
confidence range; for clarity these error bars are shown only for 1/6 randomly chosen galaxies 
of the sample.
The dotted line indicates the SFR(IR) calibration of \cite{kennicutt1998}.} 
\label{fig_lir_sfr}
\end{figure}

\begin{figure}[htb]
\centering
\includegraphics[width=8.8cm]{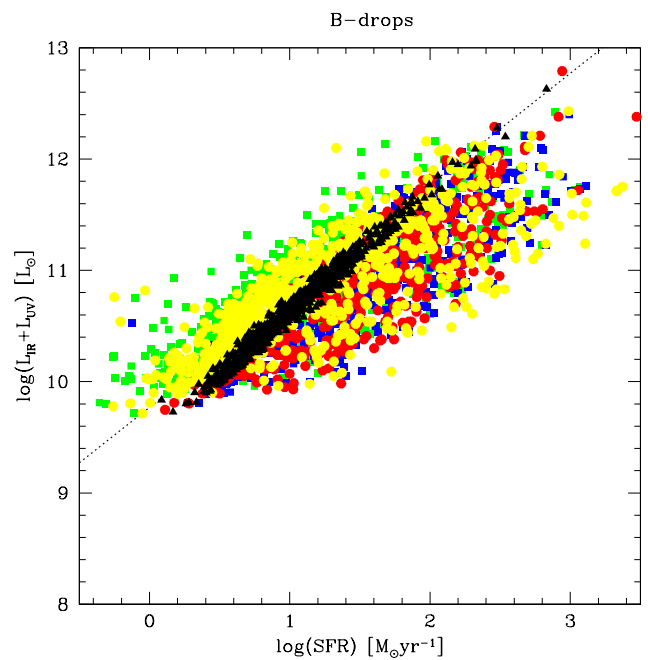}
\caption{Same as Fig.\ \protect\ref{fig_lir_sfr} for the IR+UV luminosity, which provides a better 
approximation of the bolometric luminosity.} 
\label{fig_lbol_sfr}
\end{figure}

\begin{figure}[htb]
\centering
\includegraphics[width=8.8cm]{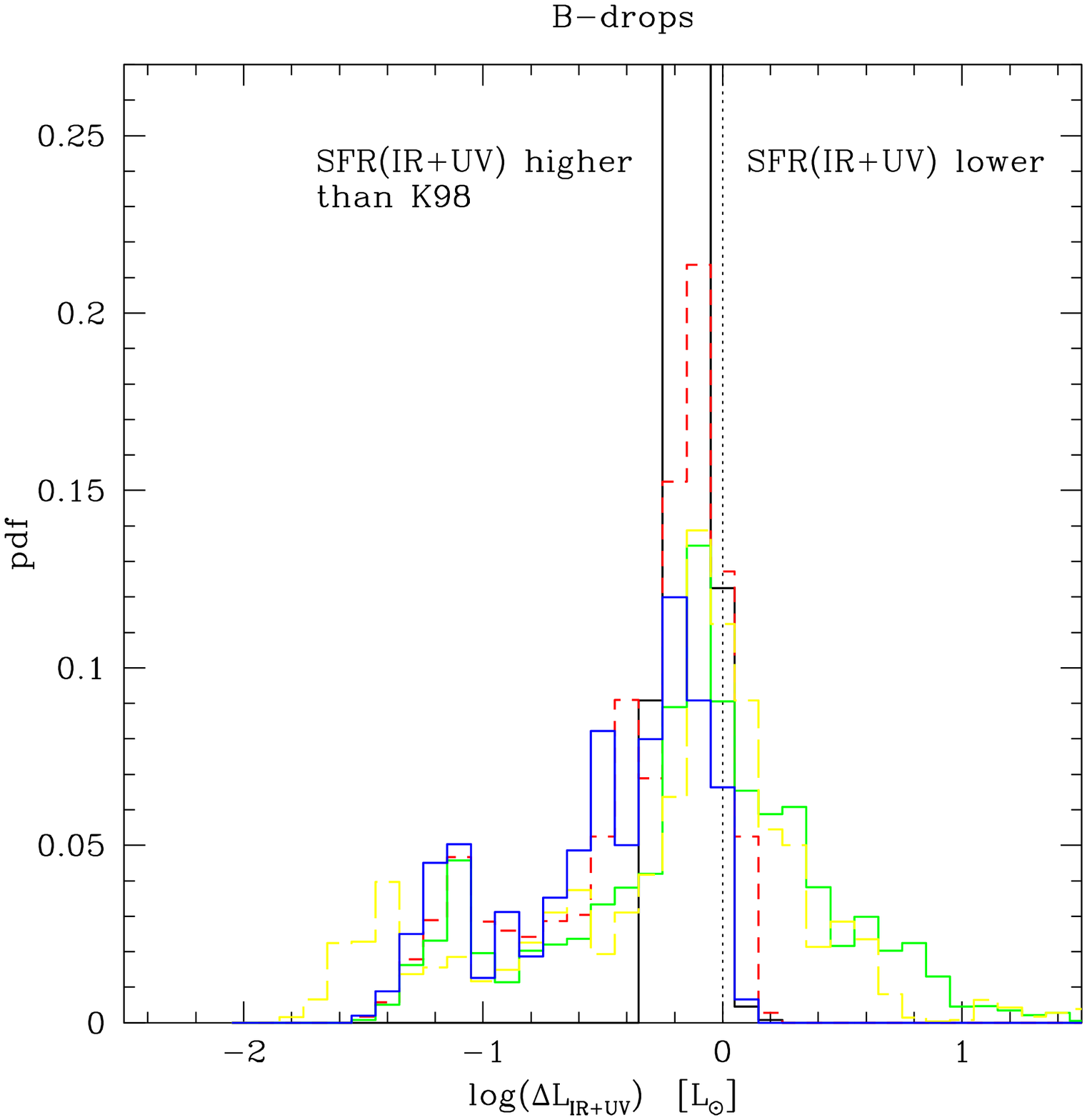}
\caption{Probability distribution function (pdf) of the difference in IR+UV luminosity, $\Delta \liruv$, between the prediction from our SED fits
for the entire B-drop sample and the luminosity derived from the SFR from relation SFR(Lbol) calibration of \citet{kennicutt1998}. 
Model A: green solid line, B: black solid, C: blue solid, D: red dashed, E: yellow dashed.}
\label{fig_k98compare}
\end{figure}

In Figs.\ \ref{fig_lir_sfr} and \ref{fig_lbol_sfr} we show the predicted IR and IR+UV luminosity respectively 
as a function of the (current) SFR determined from fitting our models to the sample of $z \sim 4$ galaxies. 
The ``standard'' SFR(IR) calibration given by \cite{kennicutt1998} -- based on the bolometric luminosity -- is also 
indicated for comparison.
Overall quite large deviations from the simple calibration are obtained for \lir---SFR, and smaller deviations for
$(\lir+\luv)$--SFR. 
As expected model fits assuming constant SF (model B) agree very well with the Kennicutt relation when the IR+UV
luminosity, providing a very good measure of the bolometric luminosity, is used. However, for galaxies with SFR $\la$ 20--30 \msunyr,
the predicted IR luminosity is decreased with the respect to this relation. This is simply due to the fact that these 
galaxies are not found to be dusty enough to reprocess a significant fraction of their UV light into IR radiation.
This is also clearly illustrated by the large uncertainties in \lir, which can reach values down to 0 if the pdf
of \av\ does so. This effect, shown here for SFR=const, is present for all star formation histories.

More generally, for varying star formation histories, both the IR and the IR+UV luminosity predicted by model fits
are  {\em lower} than expected from the Kennicutt relation between SFR and luminosity (Figs.\ \ref{fig_lir_sfr} and  \ref{fig_lbol_sfr}).
Two effects explain these differences.
First, dust extinction needs to be high enough for \lir\ to be a good tracer of the total SFR (cf.\ above).  
Second, the ages are younger than the time needed to reach the bolometric output assumed by \cite{kennicutt1998}.
Typically, this is the case for ages $t/\tau \la 1$ for declining SFHs, somewhat longer for delayed and rising star formation histories,
and $\sim$ 100 Myr for constant SF, as illustrated in Fig.\ \ref{fig_sfr_mbol}.
The second effect explains the values of $\lir+\luv$ below the canonical \lir--SFR relation. 
Added to that, the first effect further diminishes \lir\ for galaxies with lower SFR, as already explained above.
In some cases, although for a small fraction of galaxies, we find that the predicted IR or IR+UV luminosity
is {\em higher} than what is expected from the \cite{kennicutt1998} relation using the model SFR. This is found 
with declining or delayed star formation histories for galaxies where $t/\tau \gg 1$, i.e.\  where star formation
starts to decrease but longer-lived stars still contribute to the bolometric luminosity.
These objects are expected to show weak emission lines, i.e.\ they should include galaxies
from the category of "quiescent" LBGs identified by dBSS12,
and which can also be recognized by their 3.6 micron excess, if at $z \sim$ 3.8--5.

To illustrate more quantitatively the possible deviations from the ``standard" SFR calibration, we show in  Fig\  \ref{fig_k98compare}
the full probability distribution function (pdf) of the difference in IR+UV luminosity, $\Delta \liruv$, between the prediction from our SED fits
for the entire B-drop sample and the luminosity for the same SFR following the Kennicutt relation.
Negative values of $\Delta \liruv$ correspond to cases where the model SFR is higher than what would be expected 
from the standard calibration, positive values the opposite. Of course, the  pdf shows the same offsets 
already seen from the median values plotted for each galaxy in Fig.\ \ref{fig_lbol_sfr}. The median offsets from the 
Kennicutt relation are $-0.21$ dex with a distribution systematically shifted to lower luminosity (higher SFR) for the rising SFH with 
variable timescales (model D). For the declining histories (model A) the median offset is only $-0.07$ dex, since the distribution
extends to both higher and lower \liruv. For model B (SFR=const) a small shift ($-0.14$ dex) is also found, 
since \luv\ does not completely measure the total bolometric luminosity for galaxies with low SFR and low dust content.
More important, although the median offsets are relatively small, the models with variable SFHs predict deviations which can be quite high.
For example, deviations between $-0.83$ ($-0.66$) and $-0.04$ (0.46) are found within 68\% confidence for model C (A).

As already discussed, part of the deviation is due to relatively young ages found. Indeed,
imposing an age prior ($t>50$ Myr, e.g.), reduces the median offset $\log \Delta \liruv$  to --0.15 
and diminishes the scatter (68\% confidence range now from --0.38 to 0) for rising SFHs, and to a median of 
--0.03 (68\% confidence from --0.12 to 0.80) for exponentially declining histories. 
If only the UV {\em or} the IR luminosity, not both,  are available, the deviations and scatter are obviously even
larger than shown in Fig.\ \ref{fig_k98compare}, as already seen in Fig.\ \ref{fig_lir_sfr}.
For comparison, \cite{Wilkins2012} have recently discussed the SFR(UV) relation expected at high redshift from their semi-analytical models
of galaxy formation and evolution. They find essentially no deviation from the Kennicutt calibration
(when the same IMF is used) and a scatter of $\sim 20$\%. These differences compared to our results are most likely due to the fact that
semi-analytical models may not resolve relatively short time-scales and young ages for star-formation.

Of course it should be noted that the SFR shown in the above plots is not a direct observable, since it corresponds 
to the current SFR determined from the broad-band SED fits, and since the SFR derived in this manner depends 
itself on model assumptions, such as the assumed star formation history (cf. Sect.\ \ref{s_phys} and Fig.\ \ref{fig_sfr}).
Observationally the best measure of the current SFR obtained from the SED fits should be the one measured 
by hydrogen recombination lines (e.g.\ SFR(\ha)), once corrected for dust attenuation, since these emission lines 
trace best short-lived massive stars.
From the  \ha\ flux predicted by our SED models, the consistency with SFR(\ha) can indeed by verified.

The main conclusions to be drawn from Figs.\ \ref{fig_lir_sfr} and \ref{fig_lbol_sfr} is that IR or (IR+UV) SFR calibrations
assuming constant star formation over long timescales should be used with caution for the interpretation of high-$z$ star 
forming galaxies (e.g.\ LBGs), since they are susceptible to variations of the star formation history on short timescales,
for which there are good indications (see dBSS12 and this paper). In any case, the effects discussed here need
to be taken into account to properly predict the IR emission of LBGs and for comparisons with observations.
Possible problems related to the use of inconsistent SFR indicators, and implications on the SFR--mass relation
are discussed below (Sects.\ \ref{s_liruv_mstar} and \ref{s_earlier}).

\begin{figure}[tb]
\centering
\includegraphics[width=8.8cm]{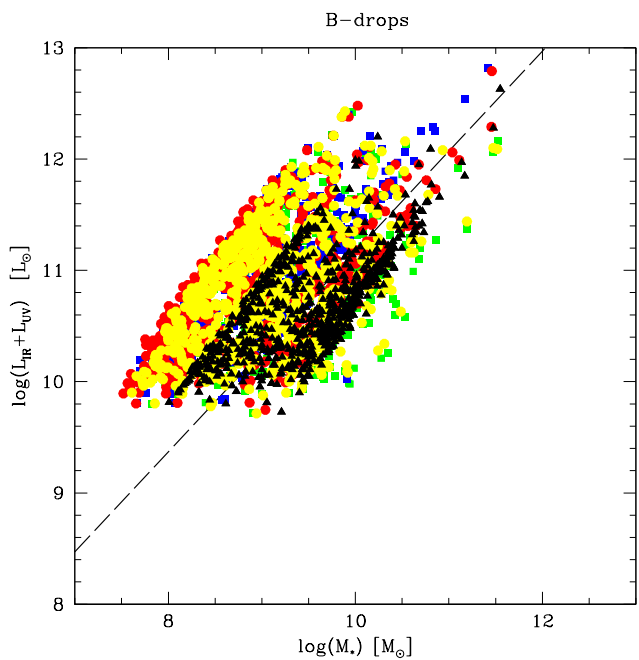}
\caption{Median UV+IR luminosity predicted consistently for the B-drop sample as a function of the median stellar mass
obtained from the SED fits for the models with different star formation histories. 
Note that the predicted scatter in luminosity is much smaller than the one in the true, underlying 
SFR shown in Fig.\ \ref{fig_mstar_sfr}, since the corresponding IR luminosity is computed consistently
with the assumed SF timescales and ages.
The dashed line indicates the expected UV+IR luminosity expected assuming the SFR--mass relation
at $z \sim 2$ from \citet{daddietal2007} and the SFR(IR) calibration of \cite{kennicutt1998}.}
\label{fig_liruv_mstar}
\end{figure}

\subsection{Implications on the SFR--mass relation}
\label{s_liruv_mstar}

Fig.\ \ref{fig_liruv_mstar} shows the SFR--mass relation with SFR transformed to direct observables,
namely the sum of the UV and IR luminosity.
Compared to the SFR--\mstar\ relation plotted in Fig.\ \ref{fig_mstar_sfr}, which shows a very large scatter (with an 
upper bound up to 2 dex above the mean relation at $z \sim 2$) for models allowing for young ages
and for variable star formation histories, the scatter in this relation is now significantly reduced, approximately 
by one order of magnitude, as judged by the upper boundary.
The reason is again due to the fact that the models with the highest current SFR/\mstar\ values correspond
to relatively young ages, where the bolometric output per SFR is lower (Fig.\ \ref{fig_sfr_mbol}). Hence 
the smaller dynamic range in bolometric luminosity, which is well approximated by $\lir+\luv$.
In other words, although the intrinsic, current SFR may show a wide dispersion, the SFR one would
infer from the UV+IR luminosity assuming ``standard" SFR conversion factors would show a much
smaller dispersion! This demonstrates that the finding of a small dispersion in the bolometric luminosity--\mstar\
plane does not exclude a high dispersion between current SFR and stellar mass, and objects with
very high values of the specific star formation rate, SFR/\mstar.

\begin{figure}[tb]
\centering
\includegraphics[width=8.8cm]{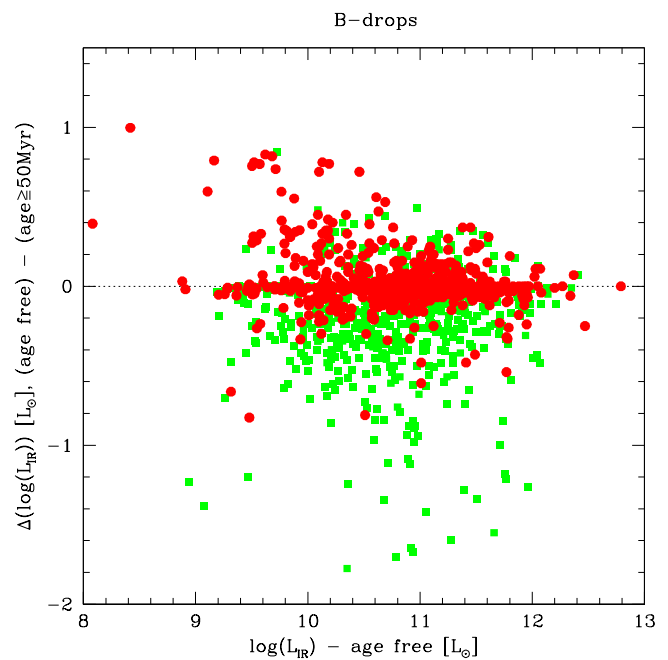}
\caption{Effect of adopting an age prior (minimum age $t>50$ Myr) on the predicted IR luminosity for declining (green, model A) 
and rising (red, model D) histories. We plot the difference in $\log(\lir)$ between the unconstrained model and the one
with age prior as a function of the latter.
On average declining SFH show lower \lir whereas rising histories predict basically the same, or even higher IR luminosities.
Rising SFHs case can thus be distinguished from constant SFR, even if age priors apply.} 
\label{fig_lbol_agemin}
\end{figure}

\subsection{Influence of the minimum age on the dust attenuation and IR luminosity}
\label{s_minage}
How does the choice of a minimum age affect the derived attenuation and hence the predicted IR luminosity?
Figure \ref{fig_lbol_agemin} shows the result of this exercise for the exponentially declining and rising SFHs
(model A and D). We find that introducing a lower limit $t_{\rm min}$ (age prior) leaves the predicted IR luminosity of LBGs 
unchanged on average for the rising star formation histories, whereas \lir\ diminishes for declining SFHs.
The basic reason for this behavior is the following: since for rising SFHs the intrinsic UV emission does not decrease when the age is 
increased, the attenuation and hence their IR luminosity remains comparable for most galaxies. In contrast,
for decreasing star formation histories an older age allows one to reproduce the SED with a lower attenuation
(cf.\ the well known age--extinction degeneracy). Thus a lower \lir, comparable to the values predicted 
for the constant SF model with the same minimum age, is expected for this case.
In other words, although the predicted \lir\ depends in general on assumptions on the  minimum age
of stellar populations, this is not the case for rising star formation histories, which consistently predict
a higher UV attenuation and hence IR luminosities than constant SF or declining histories.
A similar distinction was already seen in the SFR--mass relation (Fig.\ \ref{fig_mstar_sfr_agemin}).
Measuring \lir\ can help test/constrain these different scenarios.

\section{Predictions for ALMA}
\label{s_alma}

\begin{figure}[tb]
\centering
\includegraphics[width=8.8cm]{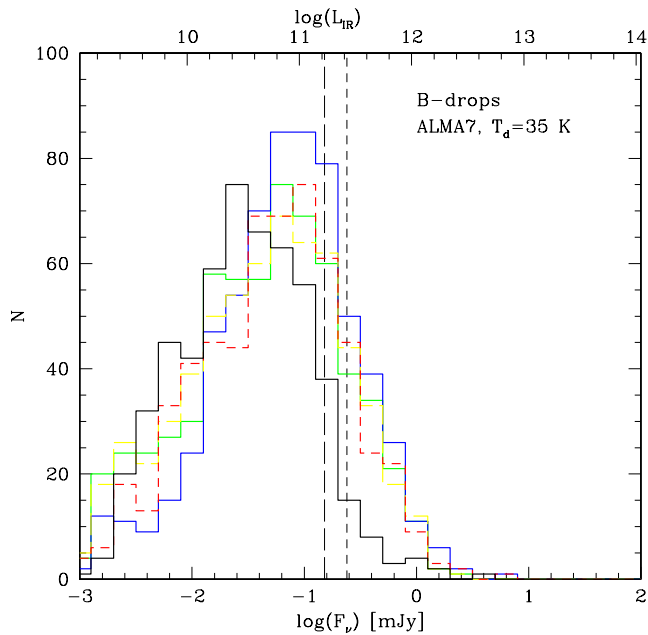}
\caption{Predicted histogram of median dust continuum fluxes in the ALMA band 7 (0.8--1.1 mm) for the B-drop sample 
from models assuming different star formation histories. The predictions are plotted for modified black bodies with $T_d=35$ K
and $\beta=2.$. Same color coding as in Fig.\ \protect\ref{fig_wha}.
The vertical lines indicate the continuum sensitivity for an integration time of 60 s with the full array (long dashed), and 
with 32 antennae available for the ALMA cycle 1 (short dashed). The upper axis indicates the corresponding IR luminosity.
Changes of the dust temperature by $\pm 10$ K correspond to a change of the flux by a factor $\sim$ 2.5--2.8 approximately.} 
\label{fig_alma7}
\end{figure}

\subsection{Continuum flux predictions}

In Fig.\ \ref{fig_alma7} we show predicted fluxes for the B-drop samples in one of the ALMA bands (band 7 covering
$\sim$ 0.8--1.1mm) available during the early ALMA cycles. As explained in Sect.\ \ref{s_models} these fluxes 
are derived from the predicted IR emission, \lir\ discussed above, with the assumption of modified black bodies with 
a typical dust temperature of $T_d=35$ K and $\beta=2$. Changes of the dust temperature by $\pm 10$ K correspond 
to a change of the flux by a factor $\sim$ 2.5--2.8 approximately. At this redshift ($z \sim 4$) the emission increases (decreases) 
with decreasing (increasing) dust temperature in this band, since the peak of the spectrum lies at shorter wavelengths
than band 7 considered here.

Fig.\ \ref{fig_alma7} clearly shows the significant flux differences predicted by models assuming different star formation
histories. If the dust properties (temperature and exponent $\beta$) do not change, the IR-mm fluxes simply reflect the 
differences in IR luminosity predicted by the models, as already discussed in Sect.\ \ref{s_ir}.
While the majority of the $z \sim 4$ LBGs show fluxes below the sensitivity limit reached with short (60 s) exposures
with the full ALMA array, approximately half of the sample should be detectable with 30min exposures.
In any case, already with the current, incomplete ALMA array the brighter part of the distribution should be detectable.
Such measurements will provide badly needed, direct measurements of the UV attenuation
of $z \sim 4$ LBGs and very useful tests/constraints on the SED models and on the star formation histories of these
distant galaxies.

When selecting galaxies by their IR luminosity, which part of the SFR--mass diagram such as Fig.\ \ref{fig_mstar_sfr} do we probe?
Naively one would expect this to translate to a simple limit in SFR. However, since SFR/\lir\ is a priori not constant, i.e.\ age dependent,
and since \lir\ also depends on the amount of dust attenuation, the limit is more complicated. 
This domain is e.g.\ illustrated for $\lir >2 \times 10^{11}$ \lsun\ by the dashed line in Fig.\  \ref{fig_mstar_sfr}.
Whereas this criterium selects all massive galaxies with SFR$\ga 35$ \msunyr, such galaxies with $M \la 6.\times10^8$ \msun\
have relatively young populations and less dust, which implies a lower IR luminosity for the same SFR, below the
IR luminosity limit. Although the detailed shape of curves of \lir=const in Fig.\ \ref{fig_mstar_sfr}  depend somewhat
on the value of \lir, their qualitative behaviour is similar.

\begin{figure}[tb]
\centering
\includegraphics[width=8.8cm]{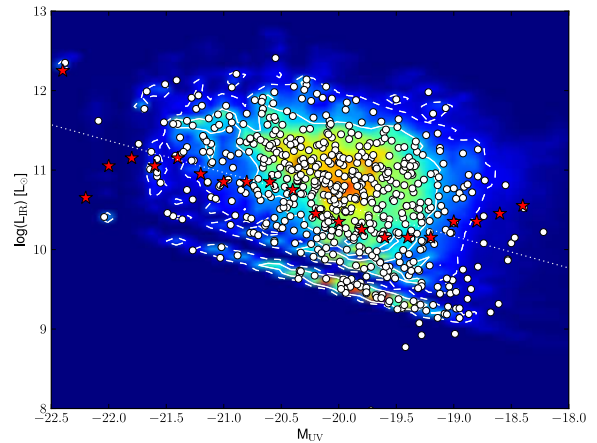}
\caption{Predicted IR luminosity as a function of the absolute UV magnitude for the B-drop sample ($z \sim 4$)
assuming declining star formation histories (model A).
Colors show the  full 2D probability distribution function (pdf) of the sample. The solid (dashed) lines encircle
the 60 (90) \% confidence range. The median values of each galaxy with a median $A_V\ge 0.05$ are shown as 
white circles. The median of the pdf in bins of UV magnitude is shown as red stars. The dotted line
shows, for comparison, the expected \lir\ when assuming the standard SFR(UV) and SFR(IR) relations of 
\cite{kennicutt1998}, and  SFR(UV)$=$SFR(IR).
} 
\label{fig_maguv_bdrop}
\end{figure}
\begin{figure}[tb]
\centering
\includegraphics[width=8.8cm]{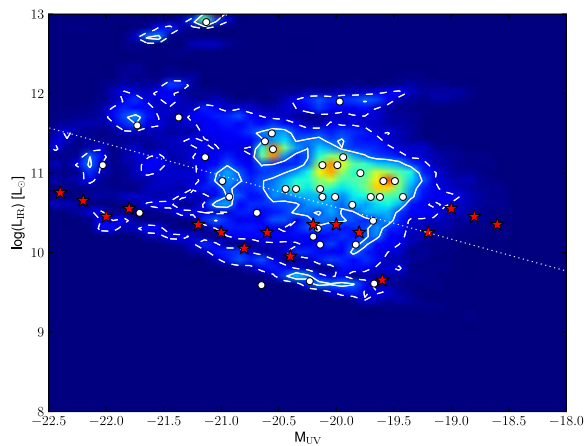}
\caption{Same as Fig.\ \protect\ref{fig_maguv_bdrop}, for the i-drop sample with 60 galaxies at higher redshift ($z \sim 6$).
Note the significant decrease in the median \lir\--$M_{\rm UV}$ relation compared to $z \sim 4$, due to the decrease
of the median dust attenuation found with increasing redshift (cf.\ dBSS12).} 
\label{fig_maguv_idrop}
\end{figure}


\subsection{Evolution of the IR luminosity with redshift}

The typical predicted IR luminosities at $z \sim 4$ are shown in Fig.\ \ref{fig_maguv_bdrop}
as a function of the absolute UV magnitude  (here $M_{1500}$ for comparison with dBSS12)
for model A, i.e.\ declining star formation histories.  
The figure shows the 2D probability distribution function (pdf), the median values for each galaxy (white
circles), and the median of the pdf in bins of UV magnitude (red stars).
As mentioned earlier, the pdf also includes a tail to very low IR luminosities, set somewhat arbitrarily to
\lir\ values outside the plot, since some of the SED fits also allow for zero extinction. This explains why the 
median IR luminosity (red stars) is shifted towards lower values than the peak of the pdf.
Incidentally, at $z \sim 4$, the median \lir\ for model A is found to be quite close to the IR luminosity
predicted from the UV magnitude using the standard SFR(UV) and SFR(IR) relations of 
\cite{kennicutt1998}, and assuming SFR(UV)$=$SFR(IR), as shown by the dotted line.
On average we find that the UV brightest galaxies are also expected to be brightest in the IR,
in agreement with \citet{reddy2012,gonzalez_durham2012} and other studies.
Typical median luminosities are of the order of $\lir \sim 10^{10-11}$ \lsun\ for model A
(cf.\ Fig.\ \ref{fig_alma7}).

\begin{figure}[tb]
\centering
\includegraphics[width=8.8cm]{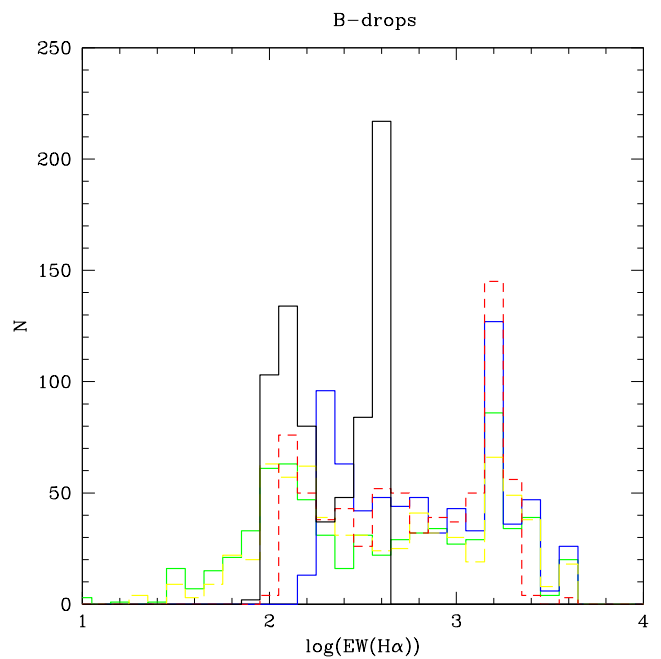}
\caption{Histogram of the predicted median (rest-frame) \ha\ equivalent widths for the $z \sim 4$ sample
from models assuming different star formation histories (Model A: green solid line;
model B: solid black; model C: solid blue; model D: red dashed; model E:  yellow dashed).} 
\label{fig_wha}
\end{figure}

In Fig.\ \ref{fig_maguv_idrop} we show the IR luminosity for the i-drop sample ($z \sim 6$)
predicted from model A. It should be noted that a higher fraction of the 60 objects of this
sample are found with a median attenuation of zero (below our minimum non-zero value of $A_V=0.05$),
implying thus a median \lir\ outside this plot. In other words the complete pdf has a secondary peak
at low values of \lir, not shown here, which explains -- as for Fig.\ \ref{fig_maguv_bdrop} --
the shift of the median of the sample (red stars) with respect to the apparent peak of the pdf.
Clearly, the predicted IR luminosity of $z \sim 6$ LBGs is lower by a factor $\sim$ 3--4
at most UV magnitudes, due to the lower dust attenuation at high redshift (see dBSS12 for 
more details). However, despite the lower reddening, the SED fits show that 
galaxies at $z \ge 6$ with dust exist quite likely, even at these high redshifts
(cf.\ dBSS12, also Schaerer \& de Barros 2010). Indeed, these studies show a clear trend
of increasing average dust attenuation with galaxy mass, at all redshifts.

\section{Predicted (rest-frame) optical emission lines}
\label{s_lines}

\begin{figure}[tb]
\centering
\includegraphics[width=8.8cm]{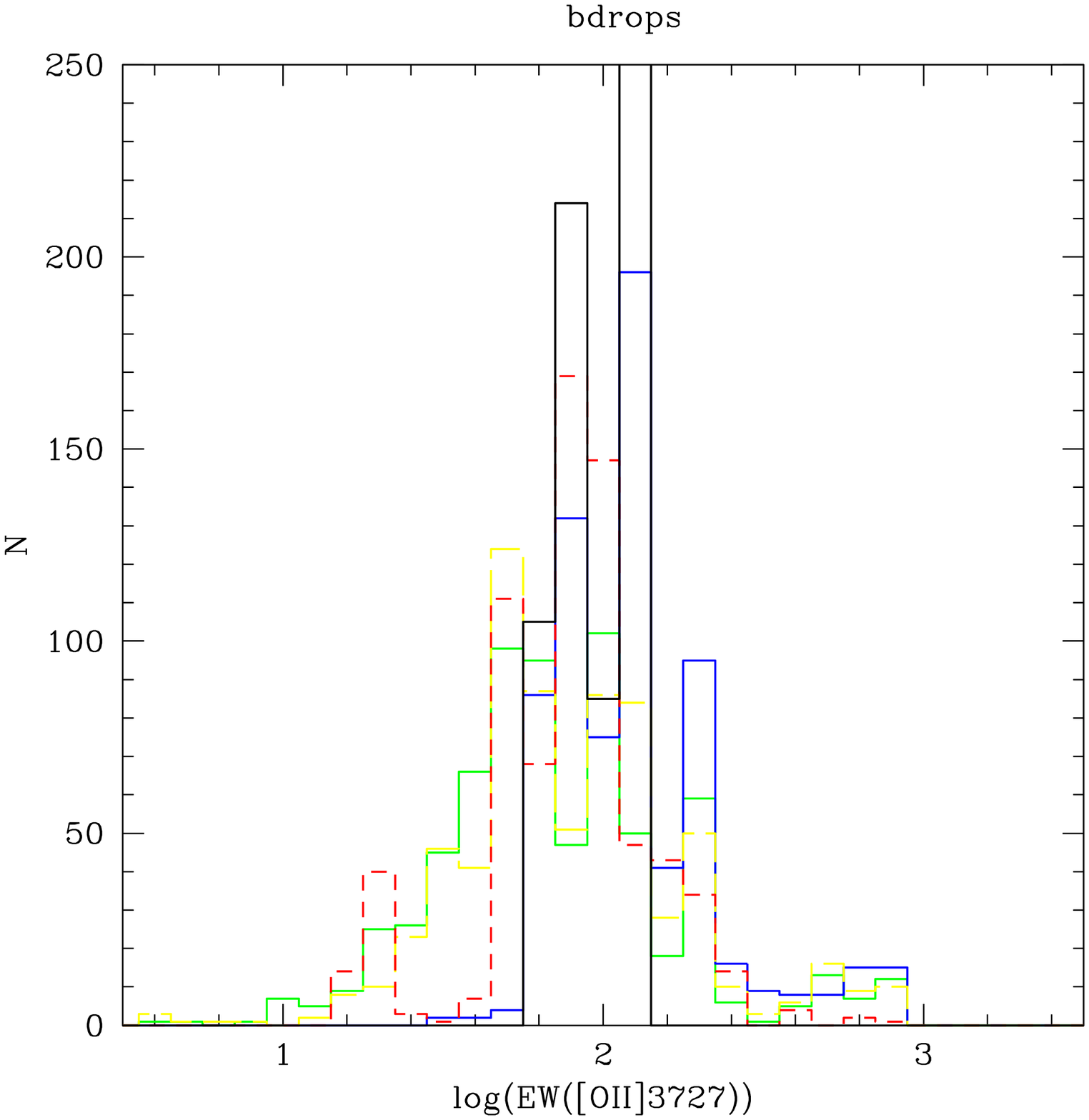}
\caption{Same as Fig.\ \protect\ref{fig_wha} for the (rest-frame) \Oii\ equivalent width.} 
\label{fig_woii}
\end{figure}

\begin{figure}[htb]
\centering
\includegraphics[width=8.8cm]{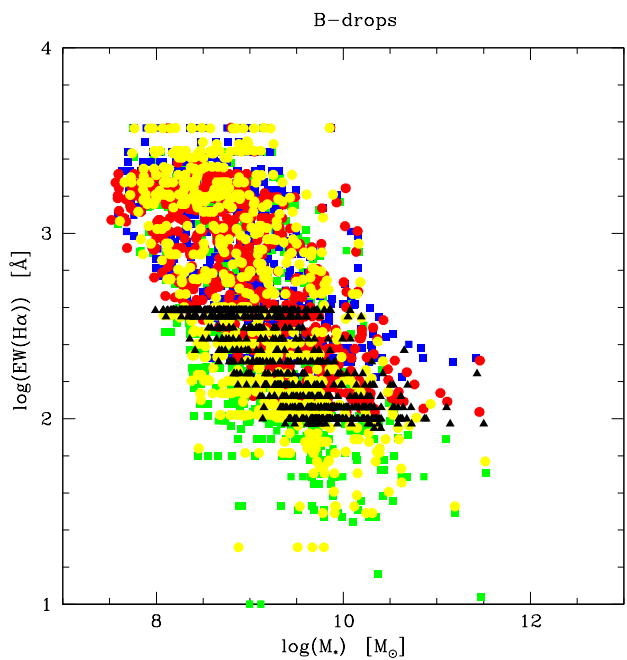}
\caption{Predicted (rest frame) \ha\ equivalent width of the B-drop galaxies as a function of the stellar mass.
Different colors indicate the different models, as in previous plots. For all SFHs an anti-correlation of
\whalpha\ with galaxy mass is expected. However, the detailed values of the equivalent width of 
each galaxy, and the range of \whalpha\ covered, are model dependent.}
\label{fig_wha_mstar}
\end{figure}

\begin{figure}[htb]
\centering
\includegraphics[width=8.8cm]{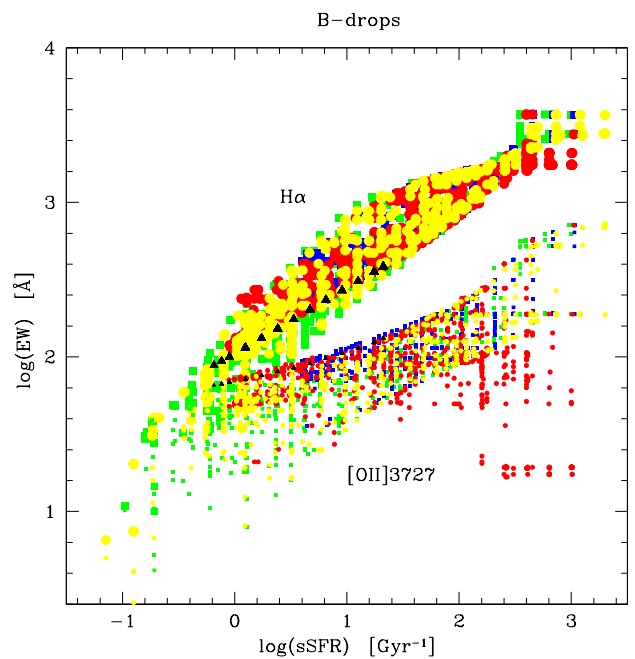}
\caption{Predicted correlation between the (rest-frame) equivalent width of \ha\ (large symbols) and the specific star formation rate.
The values shown here are derived for the sample of B-drop galaxies. However, all samples and models 
yield very similar relations.
Small symbols show the same for EW(\Oii), which traces less well the sSFR.
} 
\label{fig_ew_ssfr}
\end{figure}

Since one of the main features of our SED analysis is the treatment of nebular lines, whose effects
turns out to be not negligible for approximately two thirds of the LBGs (cf.\ dBSS12), it is obviously of interest
to examine the predicted strength of these emission lines, and to test these predictions observationally.
From the 63 lines from the (rest-frame) UV, optical and near-IR domain included in our models, we illustrate here \ha, 
among the strongest optical lines, and \Oii, the bluest optical line. Their predicted equivalent width distribution
for the B-drop sample and for the different star formation histories are shown in Figs.\ \ref{fig_wha} and \ref{fig_woii}.

The predicted \ha\ equivalent widths cover a wide range, with a maximum reaching up to $\sim 3000$ \AA\ for
some rare cases of very young galaxies. Most SFHs, except for the model with constant SFR, predict a fairly
wide distribution of equivalent widths. The rising star formation histories (models C and D) predict higher 
median equivalent widths, since the current star formation always produces Lyman continuum photons, hence
also emission lines. As expected, models with constant SFR predict a relatively narrow range of equivalent widths,
limited by a minimum \whalpha $\ga 100$ \AA, which is reached close to the maximum age of $\la 1$ Gyr.
Only models allowing for some decline of star formation (e.g.\ model A, E), can predict objects with very weak
or no emission lines, as clearly shown in Fig.\  \ref{fig_wha} and \ref{fig_woii}.
From these figures is evident that measurements of the equivalent width distribution in LBG samples can,
in principle, distinguish different star formation histories. 
In practice, however, such line measurements are currently difficult or not yet possible, due to limited 
sensitivity/access at the observed near-IR wavelengths.
For this task, recombination lines such as \ha\ should
ideally be used, as they trace directly current star formation, whereas metal lines depend more strongly
on nebular conditions (metallicity etc.). 

The \ha\ fluxes predicted for $z \sim 4$ sample are between $10^{-18}$ and $10^{-16}$  \ergscm, with
the lowest median predicted for the SFR=const model.
The predicted \Oii\ fluxes are in a similar range, but less differences are expected between different SFHs,
as already mentioned above.

Not surprisingly, we find a relatively strong (anti-)correlation of the \ha\ equivalent width with galaxy mass,
shown in Fig.\ \ref{fig_wha_mstar}. Obviously different models occupy somewhat different regions
of this plot, but in all cases do we find such a behavior. Indeed, this trend is expected, since the average
age decreases and the specific star formation rate (sSFR=SFR/\mstar) increases with decreasing
galaxy mass for all SFHs (dBSS12).
Remember, however, that selection effects and biases can affect such relations
\citep[cf.][]{stringer2011,reddy2012}.
\whalpha\ decreases also with galaxy age, but the spread in age can be very large, given various
star formation timescales. The best correlation is indeed found between the \ha\ equivalent width
and the specific SFR, shown in Fig.\ \ref{fig_ew_ssfr}.
Again, this is a natural trend, since the \ha\ equivalent width measures the ratio between
the \ha\ flux --- tracking the current SFR --- and the underlying continuum flux, which depends
on the accumulated stellar mass. We note that all models predict a very similar relation, which
also holds for all samples from $z \sim$ 3 to 6, although they occupy different regions of the
relation. Measurements of \whalpha\ (from spectroscopy or estimates from broad-band photometry)
may therefore be used to determine the sSFR of individual high-z galaxies. 
As Fig.\ \ref{fig_ew_ssfr} also shows, metal lines such as \Oii, which are not 
recombination lines, trace the sSFR less well. 
 
Our predicted equivalent widths are comparable to those found recently in emission line galaxies
at $z \sim$ 0.6 -- 2.4 for \Oii, \Oiii, and \ha\
\citep{straughnetal2009,ateketal2011,vanderweletal2011}
An anti-correlation of \whalpha\ with galaxy mass is found at all redshifts, where data is currently
available, from $z=0$ to 2--2.6, as recently discussed by \citet{fumagallietal2012}.
Furthermore, the predicted \ha\ equivalent widths of the B-drops
are in good agreement with the observations of $z \sim$ 2--2.6 LBGs by \citet{erbetal2006,mancinietal2011}
in the same mass range (typically at $\mstar \ga 10^{10}$ \msun).
Our models thus predict that this trend of increasing \ha\ equivalent width continues 
down to LBGs of lower mass.

The 47 strong \ha\ emitters identified by \cite{shimetal11} at $z \sim 4$ are included in our LBG sample,
and amply discussed in dBSS12. Their selection, based on an excess in the 3.6 \micron\ filter with
respect to 4.5 \micron, is sensitive to \ewha $\ga$ 200 \AA. Clearly, our method is able to detect/predict
weaker lines, although of course with increasing uncertainty (not shown here).
The highest inferred equivalent widths from \cite{shimetal11}, well in excess of 1000 \AA, are comparable 
to our values, as are the predicted \ha\ fluxes.
Although very rare at low redshift, galaxies with \ewha\ $>$ 500 \AA\ also exist in nearby galaxies,
as pointed out by \citet{shim&chary2012}. They are all found at masses well below $\log \mstar \la 9.5$ \msun\
and at high sSFR, quite compatible with our model predictions.
Finally our predictions are also in agreement with the recent determination of \ha\ equivalent widths
for $z \sim$ 3.8--5 LBGs with spectroscopic redshifts from \cite{Stark2012}, who find  a mean value
of $\log$\ewha$ = 2.57 \pm 0.25$ \AA.

We conclude that the emission line strengths predicted by our models from fits to the observed broad-band
photometry and the correlations found are compatible with the currently existing data and trends.
Future observations of emission lines in $z \sim$ 3--6 LBGs with existing and new ground-based facilities, and with the 
James Webb Space Telescope will hopefully provide further tests on the importance of emission lines and 
constraints on the star formation histories and specific star formation rate of these galaxies.

\begin{figure}[tb]
\centering
\includegraphics[width=8.8cm]{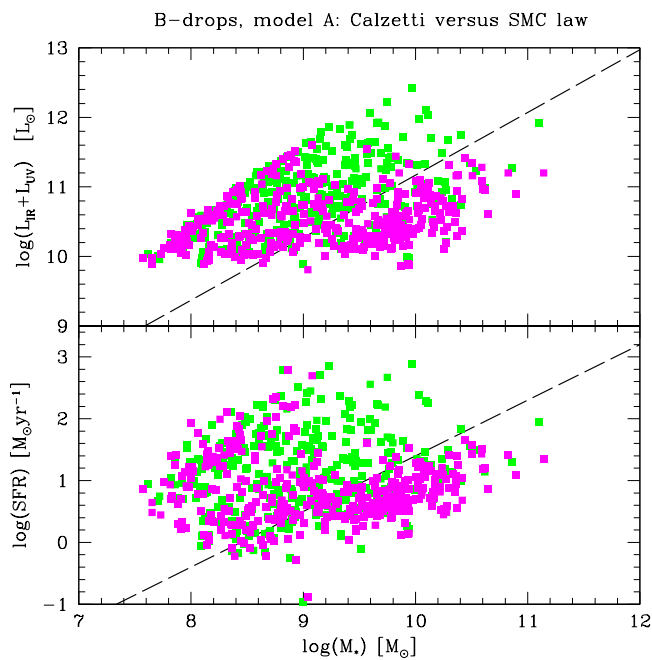}
\caption{Results/predictions from SED fits of the B-drop sample for model A (declining SFHs) assuming 
the Calzetti attenuation law (green squares) or the SMC extinction law (magenta).
{\em Top:} IR+UV luminosity as a function of the stellar mass.
{\em Bottom:} Current star formation rate versus stellar mass.} 
\label{fig_SMC_mstar}
\end{figure}\begin{figure}[htb]
\centering
\includegraphics[width=8.8cm]{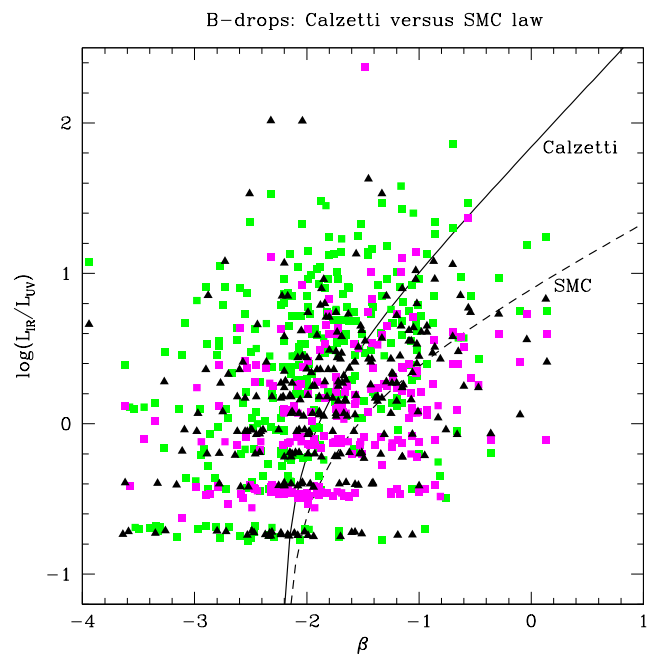}
\caption{Predicted IR/UV luminosity ratio as a function of the observed UV slope determined from SED fits of the B-drop sample 
for model A (declining SFHs) assuming the Calzetti attenuation law (green squares) or the SMC extinction law (magenta).
Black triangles show the predictions for model B (constant SFR) with the Calzetti law for comparison. The solid (dashed)
lines show the expected relation for galaxies with an intrinsic UV slope of $\beta \approx -2.2$ for the two attenuation laws.
} 
\label{fig_beta_iruv}
\end{figure}

\section{Discussion}
\label{s_discuss}
\subsection{Influence of different extinction laws}
All our SED fits have been carried out using the Calzetti attenuation law, commonly thought to be appropriate
for starburst galaxies. To examine the effect of other laws we have rerun all our fits for the B-drop sample
with the steeper attenuation/extinction law for the SMC \cite{prevotetal1984,bouchetetal1985}.
Other laws, intermediate between the relatively grey Calzetti law and the steep SMC curve, should
yield intermediate results.

Overall, we find a lower fit quality (higher reduced $\chi^2_\nu$) for the vast majority of objects using the SMC law.
However, for most of them the difference remains relatively small ($\Delta \chi^2_\nu \la 1$).
We verified that no systematic effect is found as a function of galaxy age, as could e.g.\ be expected
if young galaxies (e.g.\ $t < 100$ Myr, as judged by the age derived assuming the Calzetti law) 
would be better fit with the SMC law, as suggested e.g.\ by \citet{reddy2010,reddy2012}. 

Since the SMC law is steeper than that of Calzetti, SED fits for LBGs are generally yield a
lower UV attenuation $f_{\rm UV}$ (and even more so a lower $A_V$), older ages, and hence on average somewhat 
larger stellar masses \citep[e.g.][]{papovichetal2001,vermaetal2007,yabeetal2009,reddy2010}. 
As expected, our fits including nebular lines also yield the same average trends. The young ages
favored with the Calzetti law for some objects by the strong \ha\ excess at 3.6 \micron\ are 
in part compensated by increasing the star formation timescale $\tau$ for declining SFHs.
The resulting \ha\ equivalent width distribution (cf.\ Fig.\ \ref{fig_wha}) remains quite similar,
with the largest differences due to the assumed star formation histories.

Important differences of the SED fits with the SMC law are in particular a lower SFR and less IR emission.
The former is due both to the older age (hence lower SFR/UV, cf.\ Fig.\ \ref{fig_sfr_mbol}) and to the lower UV attenuation.
The lower attenuation implies immediately a lower IR luminosity. 
For model A (declining SFHs) the median SFR of our B-drop sample is lowered by a factor $\sim 2$, from
11.6 to 5.4 \msunyr, the median IR luminosity from $4.\times 10^{10}$ \lsun\ by a factor $\sim 5$, compared to SED fits with the Calzetti law. 
The effect of adopting the SMC law on the SFR--mass on the (\lir+\luv)--mass relations for model A are shown
in Fig.\ \ref{fig_SMC_mstar}. The top panel clearly shows the reduced IR luminosity, implying a lack of B-drop
galaxies with $\lir \ga 2.\times 10^{11}$ \lsun\ (in this sample) if the SMC law was applicable. 
A large scatter is still found in the SFR--mass relation with some objects reaching large specific SFRs, as shown
in the bottom panel. However, at masses $\ga 10^9$ \msun\ the SED fits with the SMC law predict
basically no LBGs with high specific SFR, i.e.\ above the $z \sim 2$ relation of \citet{daddietal2007}.
This shift to lower SFR values is again mostly due to age shifts, leading to ages $\ga$ 100 Myr for all
galaxies with $M \ga 10^9$ \msun. It must, however, be remembered that the model SFR shown
here represents the current SFR, which shows a wider dynamic range (spread) than the observable UV and IR 
luminosity (cf.\ Sect.\ \ref{s_liruv_mstar}).

At low redshift, and ocasionally at $z \sim$ 2--3, observations of  the IR/UV ratio and measurements
of the UV slope $\beta$ have been used to distinguish the attenuation law \citep[e.g.][]{bakeretal2001,sianaetal2008,sianaetal2009,reddy2010}.  
The so-called IRX--$\beta$ plot, based on the predicted IR luminosity
plus the observed \luv\ and $\beta$, for our $z \sim 4$ LBG sample and for the two attenuation laws
is shown in Fig.\ \ref{fig_beta_iruv}. As already mentioned above, a lower IR/UV ratio is found 
from SED fits using the SMC law. Besides an accurate measurement of the IR/UV ratio (soon feasible with ALMA), 
this method requires also a good knowledge of the intrinsic UV slope, and accurate 
measurements of $\beta$  to constrain the attenuation law. 
Indeed, as clear from this plot, UV slope measurements of individual galaxies
show a large scatter and their errors are relatively large since 
the  $\beta$ slope depends on one UV color, here  $\beta \propto 5.3 \times (i-z)$ color, following \citet{Bouwens09_beta}.  
More accurate measurements are feasible, e.g.\ using more photometric bands
or based on SED fits \citep{finkelsteinetal2011,castellanoetal2012}.
In any case, the use of the IRX--$\beta$ plot to distinguish attenuation laws also requires the knowledge
of the intrinsic UV slope, which may not be unique, since the intrinsic UV slope depends on the star formation history,
age, and metallicity of the stellar population.

Distinguishing different attenuation laws may thus not be straightforward. However, some combinations
of star formation history and attenuation law are clearly  distinguishable once IR luminosities
are measured with ALMA.  As already mentioned, our SED fits to the present data
yield in general better fits with the Calzetti law for the vast majority of objects, although the 
difference in $\chi^2_\nu$ is not very large. 

\subsection{Comparison with earlier studies}
\label{s_earlier}
We shall now briefly discuss other papers, which have presented tests for or constraints on different star formation
histories for Lyman break galaxies using also the IR as a constraint.

To the best of our knowledge, our study is the first discussing SED fits with various star formation
histories including both declining and rising parametrisations for a large number of $z \sim$ 3--6 LBGs.
At somewhat lower redshift \citet{reddy2012} have recently examined a sample of $\sim$ 300
LBGs with MIPS 24 \micron\ detections and spectroscopic redshifts $1.5 \le z \le 2.6$.
The 24 \micron\ flux is used to infer the IR luminosity, which in turn provides an independent
constraint on the SFR and dust content \citep[as discussed above and in earlier papers, cf.][]{reddy08}.
From their work \citet{reddy2012} conclude that rising star formation histories provide
a better agreement between different SFR indicators, and hence that declining star formation
histories may not be  accurate for typical galaxies at $z \ga 2$.

This conclusion appears in contrast to ours, since we find that declining
SFHs provide the best fits for the majority of $z \sim$ 3--6 LBGs, and that delayed
and rising star formation histories fare comparably. 
First, we have to note that based on the SED fit quality (\ki2) alone \citet{reddy2012} also find quite similar
results between their different star formation histories. They then use additional constraints 
(mid-IR observations) to examine SFR indicators for their consistency. 
The principle of their method is exactly what we're promoting to use. However, such observations are 
not yet available for LBGs at higher redshift. We therefore think that it is premature to 
draw general conclusions on the star formation histories of LBGs at $z \ge 3$ from the dataset
of bright $z \sim 2$ galaxies studied by  \citet{reddy2012}.

We note that in our analysis of $z \sim$ 3--6 LBGs with declining star formation histories
we find values of $t/\tau < 1$ for the majority of objects, in agreement with the 
observational arguments mentioned by \citet{reddy2012}, and in contrast to their finding
of  $t/\tau > 1$ for the majority of $z \sim 2$ sample.
Furthermore, all the currently available constraints for $z \sim$ 3--6 LBGs can be well reproduced
with variable (i.e.\ non-constant) SFHs, whether exponentially declining or rising (cf.\ also dBSS12). 
In any case, stellar populations/SFHs of LBGs at higher redshift may be different, and from the result of 
\citet{reddy2012} at $z \sim$ 1.6--2.4 one cannot generalize that exponentially declining SF histories are not 
appropriate for LBGs at $z \ge 3$, since additional observational constraints -- e.g.\ from the IR --
are lacking at these redshifts. It is therefore important to obtain direct measurements of the IR luminosity at 
higher redshifts to try distinguish different star formation histories of high-$z$ galaxies.

Another issue concerns the small subsample of ``young" LBGs discussed by \citet{reddy2012}.
Indeed, for galaxies classified as ``young" ($t<100$ Myr) according to their SED fits with the 
Calzetti law, \citet{reddy2012} find that the SFR inferred from the fits, SFR(SED), is systematically
larger than the SFR(IR+UV) using the standard calibrations
of \cite{kennicutt1998}. From this they conclude that there is an apparent conflict with young ages,
which they propose to resolve by invoking a different, steeper attenuation law for young objects.
We are not entirely convinced by the reality of this conflict, and hence by the proposed solution.
Indeed, although they note that a young age is not compatible with 
assumptions made for the standard UV--SFR conversion, they neglect the fact that both IR and
UV conversions depend on the SFH and age in a similar way (see their Fig.\ 25 and Fig.\ \ref{fig_sfr_mbol}),
modifying the (UV+IR)---SFR conversion upward by typically twice the amount they state.
The effect they find is clearly shown to be predicted for our $z \sim 4$ sample, as illustrated in Fig.\ \ref{fig_lbol_sfr},
where we note that for the majority of objects fitted with unconstrained ages and found to be younger than 100 Myr from our models,
the SFR(SED) is larger than the SFR(IR+UV)  obtained from an inconsistent application of the 
standard calibrations. Offsets by up to a factor $\sim$ 10 from the standard Kennicutt relation
can be found (cf.\ Fig.\ \ref{fig_k98compare}).
Whether this can quantitatively solve the (sometimes large) discrepancies found by  \citet{reddy2012} remains, 
however, to be seen.

Other comparisons with \citet{reddy2012} regarding constant and rising star formation histories
have already been discussed in dBSS12, and shall not be repeated here.

\subsection{Derived ages and implications for the galaxy populations}
\label{s_age}
At first sight the ages derived from our models and the non-evolution of the median mass
at a given UV magnitude (cf.\ dBSS12) imply that a fraction/most of the LBGs 
observed in a given sample (e.g.\ at $z \sim 4$) would not have been seen at earlier
epochs (e.g.\ in the $z \sim 5$ and 6 samples). This seems e.g.\ opposite 
to the finding of \citet{papovichetal2011}, who argue that LBGs are visible at multiple epochs.
Furthermore our young ages and age spread are also related to the scatter 
we find in the SFR-mass relation and to the relatively high sSFR we find at
high redshift (dBSS12), with important implications, if true.
This raises two immediate questions: First, are our derived ages correct and physically plausible?
And second,  can our results be reconciled with the picture of LBG populations with 
star formation histories rising on average?

It must be recognized that stellar population ages and uncertainties are strongly model dependent,
and that both the age uncertainties for individual galaxies and the age spread 
around the median for a galaxy sample are fairly large.
Ages obtained with models including nebular emission are generally younger
than without (cf.\ dBSS12). For the majority of galaxies the derived age is found 
to be higher than the dynamical time, and our models yield galaxy ages which increase
both with galaxy mass and with decreasing redshift, as expected from the progressive
build-up of galaxies. For a fraction of up to 20\% of the LBGs our models give
ages shorter than the typical dynamical time, estimated between 40 Myr at $z\sim 3$ 
and $\sim 20$ Myr at $z\sim 6$ (cf.\ dBSS12), which may be unphysical.
Imposing longer star formation timescales (larger $\tau$) allows one to reduce this 
fraction. However, a fraction of galaxies in all our samples is best fit with young
ages. Incidentally, these galaxies turn out to be among the least massive, 
 which at $z \sim 4$ are characterized by a very blue  (3.6-4.5) micron color,
 indicative of strong \ha\ emission. Besides this potential problem with young ages 
 for a small ($\ga$ 20\%) fraction of LBGs, we should be aware that the
 SED fits may be mostly sensitive to the most recent SF event, due to the well
 known ``outshining" effect \citep[cf.][]{wuytsetal2009,marastonetal2010}. For this reason age
 and also the galaxy mass may be underestimated, although it remains difficult
 to establish by how much.
 
 Our finding of relatively short SF timescales and young ages, and the non-evolution
 of the mass--UV magnitude relation from $z \sim 5$ to 3 (dBSS12), favours episodic star formation,
 in agreement with the findings of \citet{stark09}. If successive SF events
 take e.g.\ place with increasing strength and emission from the most recent event dominates 
 the SEDs at most wavelengths, this scenario could be reconciled with the picture of 
 slowly rising star formation histories of LBGs inferred from the evolution of the
 UV luminosity function by \citet{papovichetal2011}. Whether this works out quantitatively
 remains, however, to be seen.
 In any case the observational tests proposed here, using dust emission in the IR and optical 
 emission lines, will also help to answer these questions.
 
\subsection{Possible caveats}
\label{s_caveats}
Before closing this section, we wish to remind the reader that our SED fitting tool
suffers from the same possible limitations as other models, including e.g.\ 
the unknown star formation histories of galaxies (modeled here with simple
parametrisation), the question of multiple populations and the ``outshining" problem
due to young populations, uncertainties related to the extinction (which law? 
a unique attenuation for all stars? etc.), uncertainties in the stellar initial mass function,
and of course also uncertainties in stellar evolution and atmosphere models.
Some of these issues are e.g.\ addressed by 
\citet{marastonetal2010,wuytsetal2009,charlot2001,walcheretal2011,levesqueetal2012,schaerer2012}.
The reader should also be aware that the timescale of UV and IR emission, 
which is fairly similar in our models and in the ``standard" star formation rate calibrations, 
could be different in more complex situations, e.g.\ in cases of time-dependent attenuation, multiple populations
or alike \citep[e.g. the models of ][]{dacunha2008}.
However, such models include additional free parameters, and have, to the best of our knowledge, 
not yet been used to fit  distant galaxies. For simplicity we have therefore fitted the SEDs with single stellar populations 
and a single extinction/attenuation law.

In this respect our results should of course be considered as ``differential", i.e.\ with respect
to other models making the same/similar assumptions, but neglecting the effects of nebular emission
and of variable star formation histories, which are explored here systematically for the first time for a large sample 
of Lyman break galaxies.
Concerning nebular emission our models make some simple assumptions, which may also
influence the results. Our main assumptions are no loss of ionizing photons (i.e.\ no
Lyman continuum escape, no dust inside the \hii\ regions etc.), and the same attenuation
for the stellar continuum and nebular lines.
Both assumptions maximize to some extent the possible contribution of nebular emission,
and could therefore lead to an overestimate of the effects of nebular emission on the 
derived physical parameters. The results would then be intermediate between the present case
with nebular emission and the models without, discussed in depth in dBSS12.
Exploring variations of the above assumptions may be interesting and worth pursuing in 
future studies. 

In any case, it is clear that for a majority (approximately two thirds) of LBGs the inclusion 
of nebular effects improves the quality of the SED fits, and that models without nebular lines
cannot reproduce the observed excess in certain photometric bands (most prominently 
the 3.6 \micron\ excess for galaxies with redshift between 3.8 and 5). For the bulk of the LBGs
our models should therefore provide a better description than previous models neglecting
nebular emission, despite the simplifying assumptions made.
To progress further on these questions additional observational constraints, such
as direct measurements of IR luminosities, medium-or narrow-band photometry, or 
measurements of emission lines should be of considerable help.
The predictions presented here should also provide a useful base for future comparisons
and tests of the SED models.
In parallel, our models will also be tested against observations at lower redshift
where IR and emission line measurements are available for some galaxies.

\subsection{IR-to-mm observations to determine the UV attenuation in LBGs and constrain
their SFHs}

One of main objectives of this paper has been to show how the assumption of 
different star formation histories for SED fits of Lyman break galaxies at $z \ga 3$ affects
their derived physical parameters, and in particular their inferred dust extinction.
Independent, observational data is now needed to determine dust extinction.
Three ``classical" measurements are normally used to do this:
the UV slope, a H recombination line ratio (e.g.\ the so-called Balmer-decrement
for optical lines), or the IR/UV ratio.
Since the UV slope has already been used for our SED models, and H emission lines
beyond \lya\ are currently difficult or impossible to obtain for $z > 2.3$, measurements
of the IR/UV ratio are needed.

The deepest IR/sub-mm images available to date, taken with the Herschel satellite in the GOODS-S field, 
have detected few individual, very bright ($\lir > 10^{12}$ \lsun) LBGs out to $ z\sim$ 2.5--3
\citep[e.g.][]{elbaz2011,burgarella2011,reddy2012}.
Stacking techniques have so far
been necessary to detect fainter, more typical galaxies, and to try extend these studies to higher redshift
\citep{magdis2010,rigopoulou2010,reddy2012_herschel}.
At $z \sim 4$, for example, the expected IR luminosity of a typical LBG with $M^\star_{\rm UV} \approx -21$
is $\lir < 10^{11}$ \lsun\ (cf.\ Fig.\ \ref{fig_maguv_bdrop}), clearly beyond the reach of Herschel.
However, detections of individual LBGs at these redshift are now becoming possible with ALMA.
Waiting for such new observations we have recently analyzed stacks of the deepest
IR/sub-mm data for the GOODS-S field, to examine whether some of the LBGs studied here
are detected or whether they are detectable in stacks, and whether all of the star formation
histories are compatible with the present data. A publication on this work is in preparation.

Other more indirect tests of the different IR luminosities predicted by the models discussed
here could be carried out using IR galaxy counts and IR background measurement. Such tests
will be carried out in the future, but are, however, clearly beyond the scope of the present work.
}

\section{Summary and conclusions}
\label{s_conclude}
Following up on our earlier detailed study \citep{dBSS12} of a large sample of LBGs from redshift $z \sim$ 3 to 6
located in the GOODS-South field,
using for the first time an SED fitting tool including the effects of nebular emission on the 
synthetic photometry, we have examined the impact of different star formation histories (SFHs) on the 
 derived physical parameters of these galaxies, on the SFR--mass relation, on different SFR indicators
 (UV and IR), on the expected dust extinction and the corresponding IR luminosity, and on the 
 expected strengths of emission lines such as \ha\ and \Oii. 
 
 To do so, we have carried out SED fits for five different SFHs including exponentially rising and so-called delayed SFHs, 
 plus the three histories already considered in dBSS12 (see Table \ref{t_sfh} and Fig.\ \ref{fig_sfh}). 
 Metallicity is also treated as a free parameter,
 and we have examined the effect of two different extinction/attenuation laws (Calzetti and SMC).  
 The usual physical parameters, stellar mass, SFR, age, and attenuation
 are derived from the SED fits to the broad band photometry reaching from the U band to 8 \micron,
 using Monte Carlo simulations to derive their median values (and the detailed probability distribution
 function, generally not discussed here). 
 We have also computed consistently the predicted IR luminosities, \lir,  for all galaxies, assuming 
 energy-conservation, i.e.\ that all the radiation absorbed by dust is reemitted in the IR.
 Finally, the predicted IR luminosities have been translated to flux predictions in various IR bands,
 assuming modified black body spectra. The \lir\ predictions allow us in particular
 to examine in a consistent way the effects of variable SFHs and ages on this observable quantity,
 showing thus significant departures from results assuming inconsistent SFR(IR) or SFR(UV)
 calibrations.
  
Our main results concerning the impact of star formation histories on the physical parameters of LBGs,
exemplified to a sample of 705 LBGs at $z \sim 4$ (B-drop galaxies), can be summarized as follows
(see Sect.\ \ref{s_phys}):
\begin{itemize}
\item Compared to commonly adopted SED fits assuming constant SFR, no nebular emission and an
age prior of $t>50$ Myr, models with exponentially declining SFHs, nebular lines and no age constraint 
yield younger ages, lower stellar masses, higher current SFR, higher specific star formation rates (sSFR=SFR$/\mstar$),
and higher dust extinction ($A_V$), as already shown in \citet{dBSS12}.
Exponentially declining SFHs yield overall the best fits (in terms of \ki2 ) for the majority of LBGs.
Based on the available SED constraints it is, however, difficult to distinguish different SFHs, although 
various arguments (in particular the distribution of emission line strengths inferred from broad-band
photometry for $ z\sim$ 3.8--5 galaxies) favor clearly variable, i.e.\ non constant, histories (cf.\ dBSS12).

\item Assuming delayed star formation histories one obtains basically identical physical parameters (and fit qualities) 
as for exponentially declining SFHs.
\item Rising star formation histories with variable timescales imply generally a similar stellar masses, and comparable or somewhat higher 
dust extinction than models assuming declining SFHs. The latter leads to the highest star formation rates and to similar or higher IR luminosities
as for declining histories.
\item Overall ``standard" models assuming constant and neglecting lines predict systematically higher stellar masses, lower extinction,
lower SFR, lower IR luminosities, and more narrow range of equivalent widths for optical emission lines than all the other star formation histories
considered here.
\end{itemize}

Combining these different physical parameters we obtain the following (Sect.\ \ref{s_ir}):
\begin{itemize}
\item We find significant deviations between the derived SFR and IR luminosity from the commonly used SFR(IR) or
SFR(IR+UV) calibration of \citet{kennicutt1998}. Such differences naturally arise,
due to differences in the derived ages and in the adopted star formation histories.
In most cases (i.e.\ for most galaxies and SFHs) we find that the Kennicutt relation will underestimate the true, current 
SFR derived from the SED fits (Sect.\ \ref{s_iruv}). Consistent SED studies including also the 
IR are therefore necessary, if the SFHs and ages may differ from those assumed in standard SFR calibrations.

\item A large scatter is found in the SFR--mass relation for models with declining and delayed SFH and no age prior
\cite[cf.][]{dBSS12}. The same also hold for models with rising star formation histories.
The scatter is reduced when a minimum age (e.g.\ $t>50$ Myr) is adopted. Even in this case, models with 
rising SFHs can show a large scatter, since high SFRs are found to the high(er) extinction.

\item A large scatter in the SFR--mass relation does not necessarily imply the same scatter in the 
\lir\ (or ($\lir+\luv$)--mass relation and vice versa, when the IR luminosity is computed consistently
from the chosen SED model (i.e.\ accounting for age and SFH effects). The same also applies 
to SFR(UV)--mass diagrams where the UV luminosity and the corresponding standard SFR conversion
is used. We suggest that the true scatter in the SFR--mass relation obtained in this way 
may indeed be underestimated, if the true star formation histories are variable on relatively
short timescales. Indeed such SFH variations can reproduce more successfully 
features related to emission lines, such as the observed 3.6 \micron\ excess in $z \sim$ 4--5 LBGs. 
They may also be more relevant to at higher redshift, where the dynamical timescales decrease
with $(1+z)^{-3/2}$.
\end{itemize}

Our consistent predictions of IR luminosities (and fluxes) show that different SFHs lead to significantly
different amounts of reddening and hence to different IR/UV luminosity ratios. Measurements of 
IR luminosities of individual LBGs or statistical samples of such galaxies can be used to distinguish
different SFHs, and hence also different specific SFRs predicted by such models. ALMA observations
will thus be able to provide independent constraints on behavior of the sSFR at high redshift, and 
on the scatter in the SFR--mass relation. 

We show predictions for the IR luminosities of B-drop and i-drop galaxies for different star formation histories
and as a function of UV magnitude. The typical/median \lir\ is predicted to be $\lir \sim 10^{10 \ldots 11}$ \lsun\
for LBGs with absolute UV magnitudes of $M_{\rm UV} \sim -22$ to $-19$.

Finally we also show the predicted strengths (equivalent widths) of the \ha\ and \Oii\ emission lines.
Again, different star formation histories naturally lead to different EW distributions, which can
in principle be used to constrain the SFHs. Our models predict on average higher equivalent widths
in low mass galaxies, in agreement with currently available observations at $z<3$, and a clear 
anti-correlation of \whalpha\ with the specific SFR.

Our predictions should in particular provide new tests using IR observations with ALMA and/or
measurements of (rest-frame) optical emission lines to obtain a better insight on the star formation
histories of high redshift LBGs, on the behaviour of the SFR--mass relations and on the evolution
of the specific SFR with redshift.

\begin{acknowledgements}
We thank Mirka Dessauges-Zavadsky, Thomas Greve, and Michel Zamojski, for discussions and comments on the manuscript, and numerous 
other colleagues for stimulating and critical discussions during the last two years. 
We also thank the referee for useful comments which helped to improve the paper. 
This work is supported by the Swiss National Science Foundation.

\end{acknowledgements}

\bibliographystyle{aa}
\bibliography{references}

\end{document}